\documentclass[preprint,review,12pt]{elsarticle}
\journal{XXX}

\usepackage{amssymb}
\usepackage{amsmath,bm}
\usepackage{afterpage}
\usepackage{subcaption}
\usepackage{colortbl,booktabs}
\usepackage{tabularx}
\usepackage{threeparttable}
\usepackage{multicol,multirow}
\usepackage [font=footnotesize,labelfont=bf]{caption}
\usepackage{rotating}
\usepackage{tikz}
\usepackage{hyperref}
\usepackage[section]{placeins}
\usepackage{afterpage}

\usepackage{lineno}

\usepackage{color}
\definecolor{R}{rgb}{1, 0, 0}
\definecolor{G}{rgb}{0, 0.5, 0}
\definecolor{B}{rgb}{0, 0, 1}

\journal{Aerospace Science and Technology}
\makeatletter
\def\ps@pprintTitle{%
  \let\@oddhead\@empty
  \let\@evenhead\@empty
  \def\@oddfoot{} 
  \def\@evenfoot{} 
}
\makeatother

\begin{document}

\begin{frontmatter}

\title{{Fusion of Simulation and Experiment Data for Hypersonic Flow Field Prediction via Pre-Training and Fine-Tuning}}
\author[address1]{Yuan Jia }

\author[address1]{Guoqin Zhao}


\author[address2]{Hao Ma \corref{mycorrespondingauthor}}
\cortext[mycorrespondingauthor]{Corresponding author.}
\ead{mahao@zua.edu.cn}
\author[address1]{Xin Li}

\author[address1]{Chi Zhang}
\author[address1]{Chih-Yung Wen \corref{mycorrespondingauthor}}
\ead{chihyung.wen@polyu.edu.hk}

\address[address1]{Department of Aeronautical and Aviation Engineering, \\The Hong Kong Polytechnic University, China}
\address[address2]{School of Aerospace Engineering, Zhengzhou University of Aeronautics,\\ Zhengzhou, 450046, China}

\begin{abstract}
Accurate prediction of hypersonic flow fields over a compression ramp is critical for aerodynamic design but remains challenging due to the scarcity of experimental measurements such as velocity. This study systematically develops a data fusion framework to address this issue. In the first phase, a model trained solely on Computational Fluid Dynamics (CFD) data establishes a baseline for flow field prediction. The second phase demonstrates that enriching the training with both CFD and experimental data significantly enhances predictive accuracy: errors in pressure and density are reduced to 12.6\% and 7.4\%, respectively. This model also captures key flow features such as separation and reattachment shocks more distinctly. Physical analyses based on this improved model, including investigations into ramp angle effects and global stability analysis, confirm its utility for efficient design applications. In the third phase, a pre-trained model (using only CFD data) is successfully fine-tuned with experimental schlieren images, effectively reconstructing velocity fields and validating the transferability of the approach. This step-wise methodology demonstrates the effectiveness of combining simulation and experiment by pre-training and fine-tuning, offering a robust and efficient pathway for hypersonic flow modeling in real-world.

\end{abstract}

\begin{keyword}
Attention CNN  \sep Pre-trained model  \sep Fine-tuned model\sep Transfer learning\sep Hypersonic flow field prediction
\end{keyword}

\end{frontmatter}

%
%
\section{Introduction}
\label{sec1}
The compression corner is a common configuration in hypersonic vehicles and plays a vital role in aerospace applications \cite{yibo2022review}, making accurate characterization of hypersonic flow-field properties highly significant \cite{changtong2020research, rudy1991computation}. However, the complex Shock-Wave/Boundary-Layer Interaction (SWBLI) creates complex shock systems and separated flows \cite{gaitonde2013progress,pasquariello2017unsteady} and makes accurately measurements of the key parameters such as the velocity field particularly challenging yet crucial. The adverse pressure gradient generated by SWBLI \cite{larcheveque2023effects} deforms the boundary layer's velocity profile \cite{kokkinakis2023high,gaitonde2023dynamics}, and the underlying flow instability mechanisms can differ fundamentally from those in unseparated flows \cite{baines1994mechanism}.  As illustrated in Figure 1, flow separation occurs near the corner due to the adverse pressure gradient, forming the separation bubble. This process induces the characteristic flow structure such as the separation shock, reattachment shock and expansion waves \cite{hao2021occurrence}. Such phenomena can adversely affect vehicle performance and structural integrity \cite{nath2024alleviation}, leading to issues including reduced maneuverability, intense surface heating and loads, increased pressure loss, and detrimental structural responses caused by unsteady flow \cite{gaitonde2013progress, li2024shock, edney1968effects}. Therefore, accurately quantifying the physical properties of the hypersonic flow field in compression corners is an important research objective.
\begin{figure}[htb!]
	\centering
	\includegraphics[trim = 3cm 5cm 5cm 5cm, clip,width=0.95\textwidth]{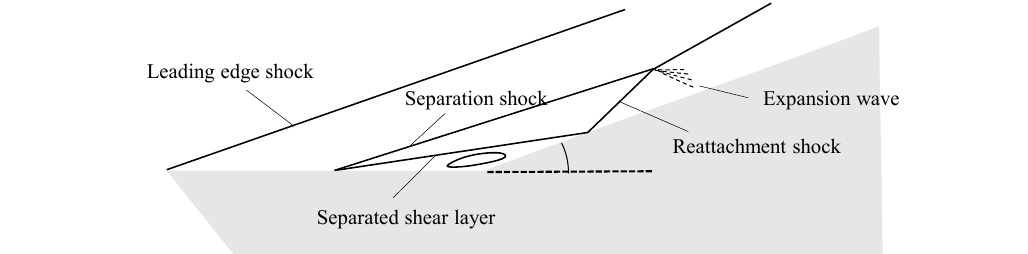}
	\caption{{Schematic of fluid structures in a compression corner.}}
	\label{figs:block-sliding-compare}
\end{figure}

Traditionally, the hypersonic flow field is obtained by the wind tunnel experiment and Computational Fluid Dynamics (CFD) simulation. However, the CFD simulation takes several hours for one case and the optimization procedures typically involve 1,000 to 100,000 iterations. With each run of the CFD solver that consumes several hours \cite{westermann2020advancing,tu2023computational,jia2021power}, the total computational time becomes intractable. On the experimental side, the acquisition of accurate velocity fields in hypersonic wind tunnels is particularly challenging. While Particle Image Velocimetry (PIV) serves as a conventional technique for velocity field measurements, its application to hypersonic flows faces significant limitations due to particle tracing fidelity issues under intense aero-thermal conditions and strong flow gradients \cite{jiang2020advances}. Given these limitations, it is of practical significance to infer velocity and pressure fields from schlieren images and sparse pressure measurements obtained experimentally. To address these challenges, the development of accurate surrogate models is essential for obtaining flow field data in a timely and computationally efficient manner. Moreover, rapid flow field reconstruction enables global stability analysis, which is crucial for structural design, as it facilitates the study of stability-dominated phenomena such as flow transition—governed by global or convective instability. In this context, leveraging schlieren imagery to extract velocity fields represents a promising and meaningful direction.

Recently, deep learning methods have gained significant traction in flow field reconstruction, with numerous successful engineering applications. Zhang et al. \cite{zhang2023improved} developed a U-Net-based Super-Resolution model (SRUnet) that effectively reconstructs high-fidelity turbulent flow fields from low-resolution data.  Duru et al. \cite{duru2022deep} proposed a deep learning approach integrated with Reynolds-averaged Navier-Stokes (RANS) simulations to efficiently predict transonic flow fields around airfoils, substantially reducing computational cost while maintaining high accuracy.  Ma et al. \cite{ma2024comprehensive} combined a surrogate model with reinforcement learning for geometric shape optimization.  However, these applications of deep learning have primarily focused on non-hypersonic flows. More recently, Li et al. \cite{li2020flow} applied a Convolutional Neural Network (CNN) to predict pressure and density gradient fields using wall surface pressure, though the model was not extended to full flow structure reconstruction. Jia et al. \cite{jia2025global} developed a fusion model for global instability prediction, yet the study provided stability assessments without accompanying physical interpretation. To address these gaps, this research employs several deep learning methods for hypersonic flow field prediction, leveraging both numerical simulation and experimental data.

Among the various neural network architectures, this study employs both CNN and Vision Transformer (ViT) for flow field prediction. To better capture boundary layer details and abrupt changes near shock waves, a coordinate transformation is applied to incorporate geometric information. Additionally, an attention mechanism is integrated into all model architectures to enhance their ability to capture critical features in the input data. Data fusion with experimental measurements further enriches the surrogate model, thereby improving its accuracy. Physical analyses, including global stability analysis, demonstrate that the results are suitable for practical engineering applications. Moreover, the fine-tuning approach shows promising potential for predicting flow fields in real-world engineering scenarios.

The remainder of this paper is organized as follows: Section 2 describes the data acquisition methods, including the experimental setup and numerical simulation approaches, as well as data preprocessing techniques. Section 3 introduces four different models based on varying neural network architectures and training datasets, along with the evaluation metrics used. Section 4 compares the performance of these models and provides further physical interpretation. Furthermore, schlieren images are utilized during fine-tuning to infer velocity and pressure fields under experimental conditions.

%
%

\section{ Dataset acquisition}
\subsection{ Experimental setup and conditions}
The data of the experiments are obtained from the Mach 4 and Mach 6  Ludwieg tube wind tunnels at the Hong Kong Polytechnic University \cite{chen2022control,zhao2024investigation}.  As illustrated in Figure 2, the facility consists of a gas storage tank,  a nozzle, a test section for model installation, and a collection tank. Figure 3 shows a schematic of the compression corner configuration, which features a flat plate of length $L =$ 100 mm.  The computational domain, bounded by the red lines, is divided into three regions: A, B, and C. Note that the schlieren image covers only region B. Along the center line of the plate, 14 Kulite pressure sensors are installed to measure the surface pressure distribution. Additional details regarding the experimental setup and measurement procedures are available in prior work \cite{zhao2024investigation}.

%
\begin{figure}[htb!]
	\centering
	\includegraphics[trim = 3cm 4cm 2.5cm 2.5cm, clip,width=0.95\textwidth]{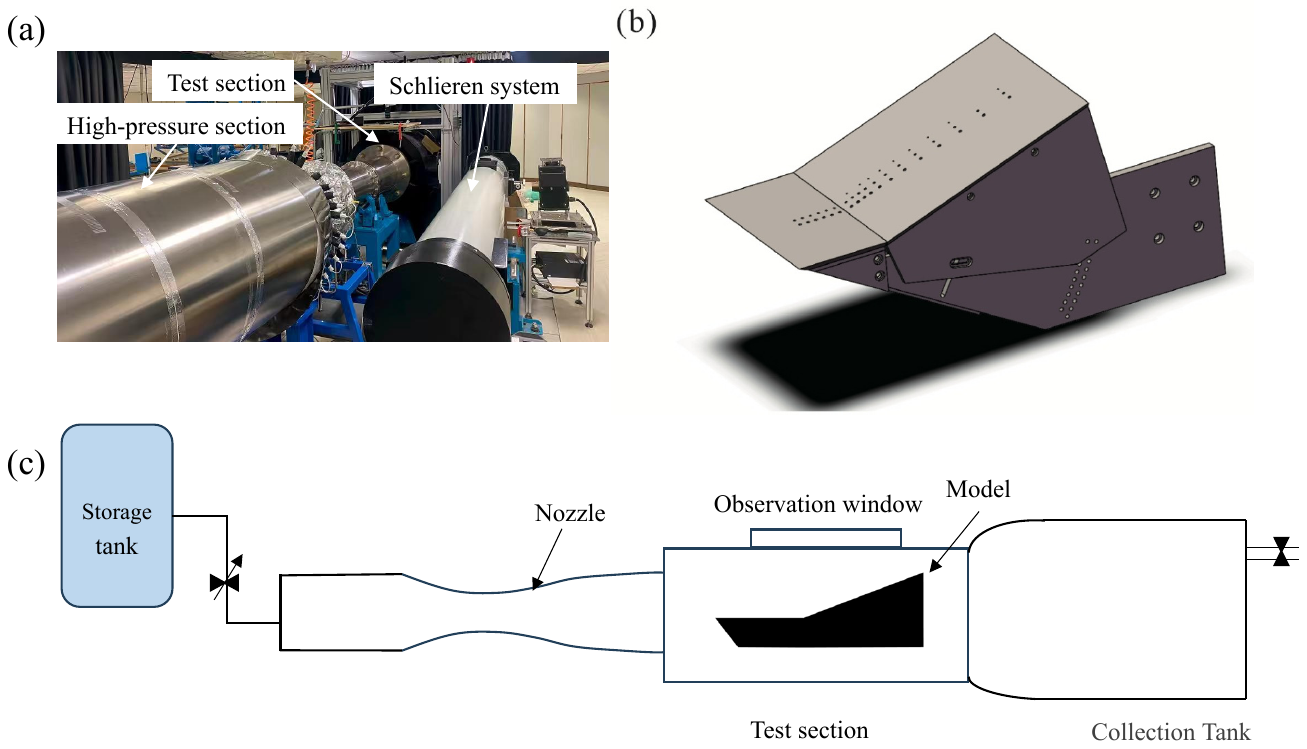}
	\caption{{(a): The experiment setup; (b): Geometry of the ramp configuration; (c): The schematic of the Ludwieg tube tunnel.}}
	\label{figs:block-sliding-compare}
\end{figure}
\begin{table}
\centering
\caption{Initial flow and geometry conditions of experiments.}
\resizebox{0.6\textwidth}{!}{%
\footnotesize 
\begin{tabular}{cccccc}
\hline
Cases   & $ Ma $            & $ Re  $            & $ T_{\infty}  $    & $ \alpha  $ & $ T_w /T_0 $ \\
\hline
Case 1 & 4 & 7220000 & 70.7 & 6° & 0.99 \\
Case 2 & 4 & 7220000 & 70.7 & 8° & 0.99 \\
Case 3 & 4 & 3660000 & 70.7 & 10° & 0.99 \\
Case 4 & 4 & 7220000 & 70.7 & 10° & 0.99 \\
Case 5 & 4 & 9190000 & 70.7 & 10° & 0.99 \\
Case 6 & 4 & 7220000 & 70.7 & 12° & 0.99 \\
Case 7 & 6 & 6370000 & 58.54 & 4° & 0.61\\
Case 8     & 6 & 6370000 & 58.54 & 6° & 0.61 \\
Case 9     & 6 & 6370000 & 58.54 & 8° & 0.61 \\
Case 10   & 6 & 6370000 & 58.54 & 10° & 0.61 \\
\hline
\end{tabular}%
}
\end{table}
The initial experiment conditions are summarized in Table 1. In this study, schlieren images are utilized for all cases, while Kulite pressure data are employed only for the $ Ma$ = 4 condition, as pressure measurements are available solely at this Mach number during training. Each schlieren image represents a temporal average of 300 instantaneous frames captured by a high-speed camera operating at 20 kHz during the steady-flow phase \cite{zhao2024investigation}.

\subsection{ Numerical methods }\label{subsec:splitting cell-linked list}
\subsubsection{ Model geometry and initial conditions }
\begin{figure}[htb!]
	\centering
	\includegraphics[trim = 5cm 5cm 10cm 5.5cm, clip,width=0.7\textwidth]{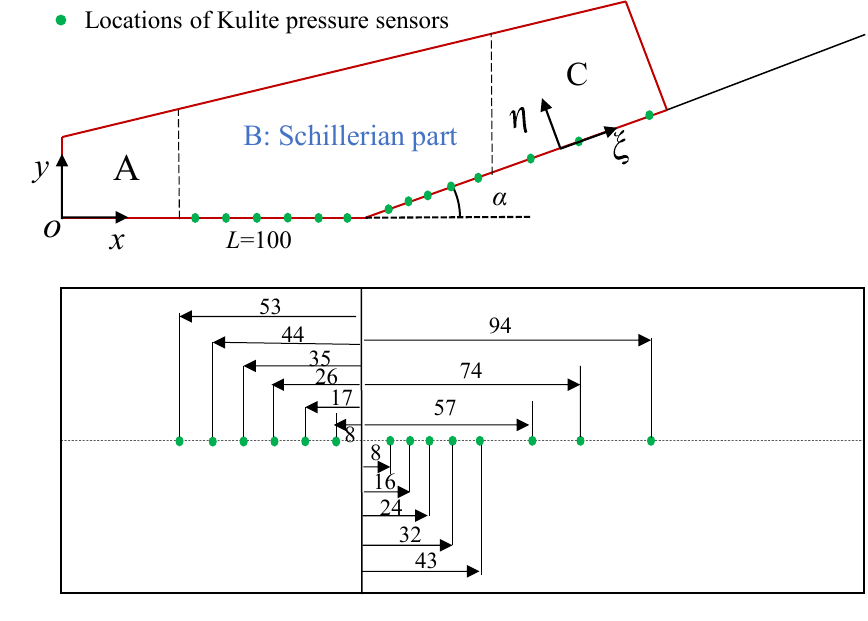}
	\caption{{The schematic of compression corner with flow simulation is composed of A, B, C three parts and the domain is enclosed by the red boundaries; The schlieren image only include B part; Green points: locations of Kulite pressure sensors. Unit: mm.}}
	\label{figs:block-sliding-compare}
\end{figure}
The compression ramp model in this study is based on the experimental configuration from the Shock Wave Laboratory at RWTH Aachen University \cite{roghelia2017experimental}, a reference geometry that has also been adopted in numerous numerical investigations \cite{hao2021occurrence,cao2023stability,hao2023response}. As shown in Figure 3, a Cartesian coordinate system $ (x,y) $ is employed with its origin at the leading edge. Additionally, a curvilinear coordinate system $ (\xi ,\eta ) $ is introduced, where $ \xi  $ represents the distance along the surface from the leading edge and $ \eta  $ denotes the normal distance from the wall. 

The initial conditions in this study are defined by five key parameters: the Mach number $ Ma$, ranging from  2 to 9.4; the Reynolds number $ Re$, from  $3.36\times 10^6$ to $5.04\times 10^6$; the freestream temperature $ T_{\infty}$,  from 100 K to 150 K; the ramp angle $ \alpha$, from  $9^\circ$ to $18^\circ$; and the wall temperature ratio $ T_w /T_0$ from 0.036 to 0.86. Here, $ T_w$ denotes the wall temperature and $ T_0$ represents the  total temperature of the freestream flow.  A total of 112 distinct combinations are generated using  Latin Hypercube Sampling (LHS) \cite{shields2016generalization} for subsequent computational analysis. 

\subsubsection{ CFD solver }
  The hypersonic flow field data for the compression ramp  applied in this study are calculated by Direct Numerical Simulations (DNS) with an in-house multi-block parallel finite-volume solver, PHAROS \cite{hao2016numerical,hao2020hypersonic}. The Navier-Stokes equations for compressible and unsteady flow can be written as:
\begin{equation}
	  \frac{\partial \textbf{\emph{U}}}{\partial \mathit{t}} + \frac{\partial \textbf{\emph{F}}_{i}}{\partial \mathit{x}_{i}}=\frac{\partial \boldsymbol{F}_{v,i}}{\partial \mathit{x}_{i}}
	\label{momentum-equation},
\end{equation}
where  $\textbf{\emph{U}} = (\rho,\rho u,\rho v,\rho e)^T$ denotes the vector of conservative variables. Here, $\rho$ stands for density; $u$ and $v$ are the flow velocities in the $x$ and $y$ directions, respectively; $e$ is the total energy per unit mass; $\textbf{\emph{F}}_{i}$ and $\boldsymbol{F}_{v,i}$ refer to the inviscid and viscous fluxes, respectively. The inviscid fluxes are discretized using the modified Steger-Warming central difference scheme \cite{maccormack2014numerical}, while the viscous fluxes are computed with the second-order central difference scheme. The boundary conditions are as follows: free-stream conditions are applied in the upper and left boundaries; an extrapolation outflow condition is applied at the exit; and a  no-slip condition is applied on the wall surface.
 
 The surface pressure coefficient $C_p$ is computed as:
 \begin{equation}
		{C}_{p}=\frac{2{p}_{w}}{{\rho }_{\infty }{u}_{\infty }^{2}},
	\end{equation}
 where $p_w$ is the wall pressure, ${\rho }_{\infty }$, and ${u}_{\infty }$ correspond to the freestream density and freestream velocity, respectively. Following the grid dependence study, a computational grid of $  1200\times  400  $ points in the streamwise and wall-normal directions is selected for numerical simulations using PHAROS. The grid has a surface-normal spacing of $  1\times  10^{-7} $ m. The grid configuration is consistent with prior research \cite{hao2021occurrence} and has been validated against experimental data. 

Figure 4 compares the pressure coefficient $C_p$ from DNS and experimental measurements for case 1 with $ Ma $ = 4,  $ Re  = 7.22\times 10^6$,  $ \alpha  $ = 6°, $ T_{\infty}$ = 70.7 K and  $ T_w /T_0 $ = 0.99.  It can be observed that for a moderate ramp angle ($\alpha$ = 6°), the DNS results match well with the experimental data.

\begin{figure}[htb!]
	\centering
	\includegraphics[trim = 0cm 13cm 0cm 0cm, clip,width=0.8\textwidth]{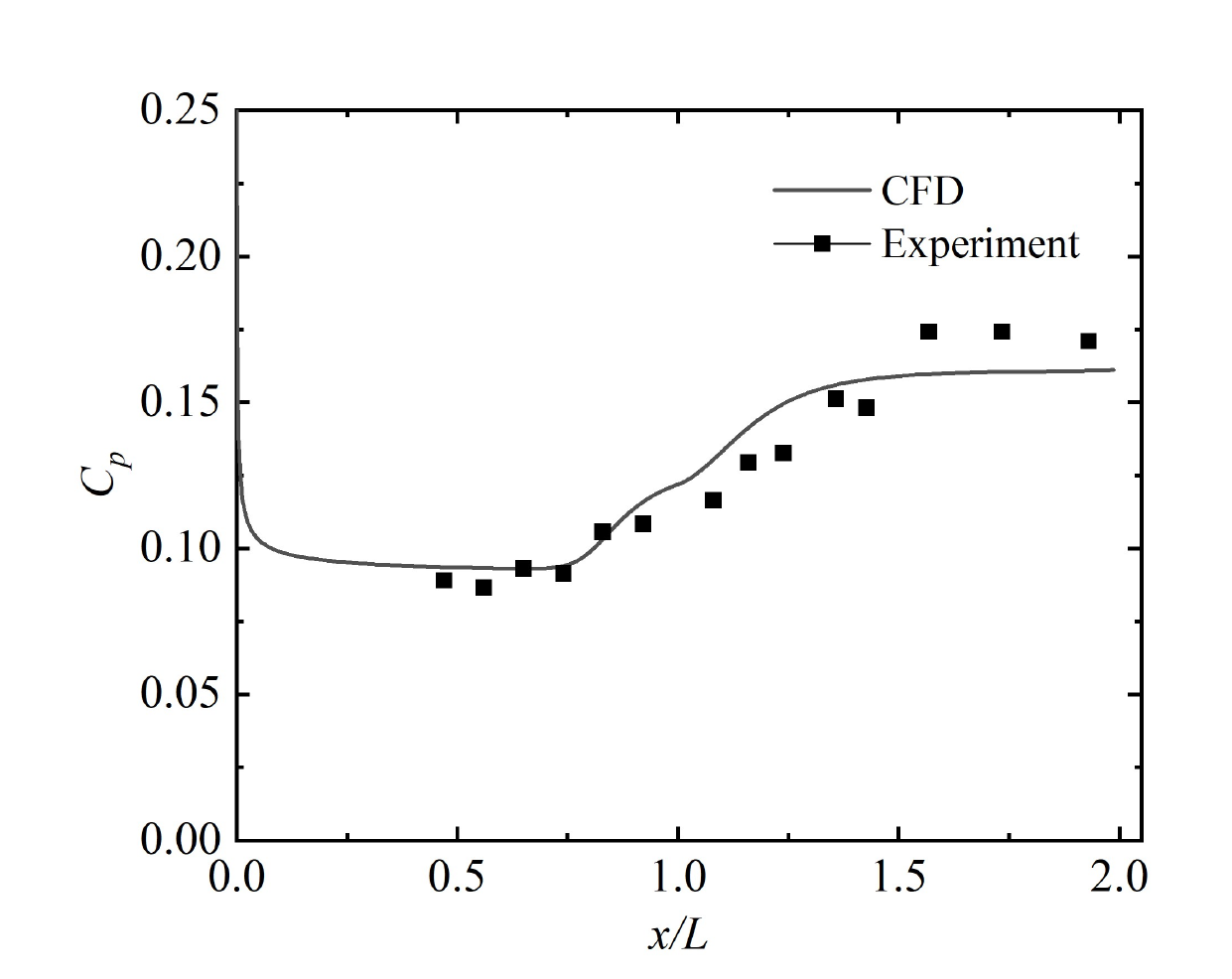}
	\caption{{Surface pressure coefficient comparison between DNS and experiment results of case 1, with $ Ma $ = 4,  $ Re   = 7.22\times 10^6$,  $ \alpha  $ = 6°, $ T_{\infty}$ = 70.7 K and  $ T_w /T_0 $ = 0.99.}}
	\label{figs:block-sliding-compare}
\end{figure}
\subsubsection{ Global stability analysis }
During the global stability process, the vector $\textbf{\emph{U}}$ can be decomposed into a two-dimensional base flow and a three-dimensional time-dependent perturbation as:
 \begin{equation}
	 \boldsymbol{U}(x, y, z, t)=\overline{\boldsymbol{U}}(x, y)+\boldsymbol{U}'(x, y, z, t),
\end{equation}
where the overbar and prime denote the base flow and perturbation components, respectively. Substituting Eq.(3) into Eq.(1) and neglecting higher-order terms yields the Linearized Navier–Stokes equations (LNS).
The perturbation $\textbf{\emph{U}}'$  is further assumed to take the modal form:
 \begin{equation}
	 \boldsymbol{U}^{\prime}(x, y, z, t)=\hat{\boldsymbol{U}}(x, y) \exp \left(\mathrm{i} \frac{2 \pi}{\lambda} z -\mathrm{i}( \omega_{r}+i\omega_{i}) t\right),
\end{equation}
where $\hat{\boldsymbol{U}}$ is the 2-D eigenfunction, $\lambda$ is the spanwise wavelength, $w_r$ is the angular frequency, and $w_i$ is the growth rate. Substituting Eq. (4) into the LNS leads to an eigenvalue problem:
 \begin{equation}
	 \boldsymbol{A}(\lambda)\hat{\boldsymbol{U}}=-( \omega_{r}+i\omega_{i})\hat{\boldsymbol{U}},
\end{equation}
where $\boldsymbol{A}$ is the global Jacobian matrix. To assess global instability, the eigenvalue problem is solved for a given parameter $\lambda$ using the shift-invert technique. This involves computing the eigenvalues with ARPACK's implicitly restarted Arnoldi method \cite{lehoucq1996arpack}, where the required matrix inversion is efficiently handled by the SuperLU library's lower-upper decomposition \cite{superlu1997superlu}. The flow is considered globally unstable if any eigenvalue exhibits a positive imaginary part.



The model is trained on 89 datasets, with 13 datasets for validation and 10 for testing.

\subsection{ Dataset preprocessing }\label{subsec:splitting cell-linked list}
A one-to-one coordinate mapping is applied from the physical space $ (x,y) $ to the computational curvilinear coordinate system $ (\xi ,\eta ) $ along the mesh directions. The detailed transformation is defined as follows:

\begin{equation}
	  \begin{bmatrix}x\\ y
\end{bmatrix}_{\hat{i},\hat{j}}=\begin{bmatrix} x_0\\ y_0
\end{bmatrix}_{\hat{i},0} +\int\begin{bmatrix}
 dx\\ dy
\end{bmatrix}
	\label{momentum-equation},
\end{equation}
\begin{equation}
	  \begin{bmatrix}dx\\ dy
\end{bmatrix}=\begin{bmatrix} x_{\xi}  & {x}_\eta\\ {y}_{\xi} & {y}_\eta
\end{bmatrix} \begin{bmatrix}
 d\xi\\ d\eta
\end{bmatrix}
	\label{momentum-equation}.
\end{equation}

In this equation, $x_0 $ and $y_0$ denote the coordinate values defining the geometry of the compression ramp. The indices $\hat{i}$ and $\hat{j}$ correspond to grid nodes in the streamwise and normal directions, respectively. The computational coordinates $\xi$ and $\eta$ are normalized as follows:
\begin{equation}
	  \xi=\frac{(\hat{i}-1)}{(\hat{i}_{max} -1)}, \eta=\frac{(\hat{j}-1)}{(\hat{j}_{max} -1)},
\end{equation}
where $\hat{i}_{max}$ and $\hat{j}_{max}$ represent  the total number of grid nodes in the streamwise and normal directions. The transformation matrix $\hat{T}$ is then expressed as:
\begin{equation}
	  \hat{T}=\begin{bmatrix} {\xi}_x  & {\xi}_y\\ \eta_x & \eta_y
\end{bmatrix} 
	\label{momentum-equation}.
\end{equation}

 Following the coordinate transformation, a supplementary set of six input parameters is obtained: $x_0$, $y_0$, ${\xi}_x$, ${\xi}_y$, ${\eta}_x$, and ${\eta}_y$.  The input feature $ \mathcal{X} $ of the neural networks consists of  the following parameters: ($ Ma $, $ Re $, $ T_{\infty}   $, $ \alpha $, $ T_w /T_0 $, $x_0$, $y_0$, ${\xi}_x$, ${\xi}_y$, ${\eta}_x$, ${\eta}_y$).

For the experiment data, the schlieren image still contains considerable extraneous information despite averaging. Therefore, a series of preprocessing steps is applied to remove irrelevant regions and enhance the visibility of flow structures. The preprocessed image has dimensions of $ 1024 \times  468$ pixels, with each pixel corresponding to $1.05\times 10^{-4} $ m.  Figure 5 shows the schlieren image before and after preprocessing for case 1, with $ Ma $ = 4,  $ Re   = 7.22\times 10^6$,  $ \alpha  $ = 6°, $ T_{\infty}$ = 70.7 K and  $ T_w /T_0 $ = 0.99. The schlieren imaging technique effectively visualizes the density gradient in the $y$ direction.
\begin{figure}[htb!]
	\centering
	\includegraphics[trim = 0cm 0cm 3cm 0cm, clip,width=0.99\textwidth]{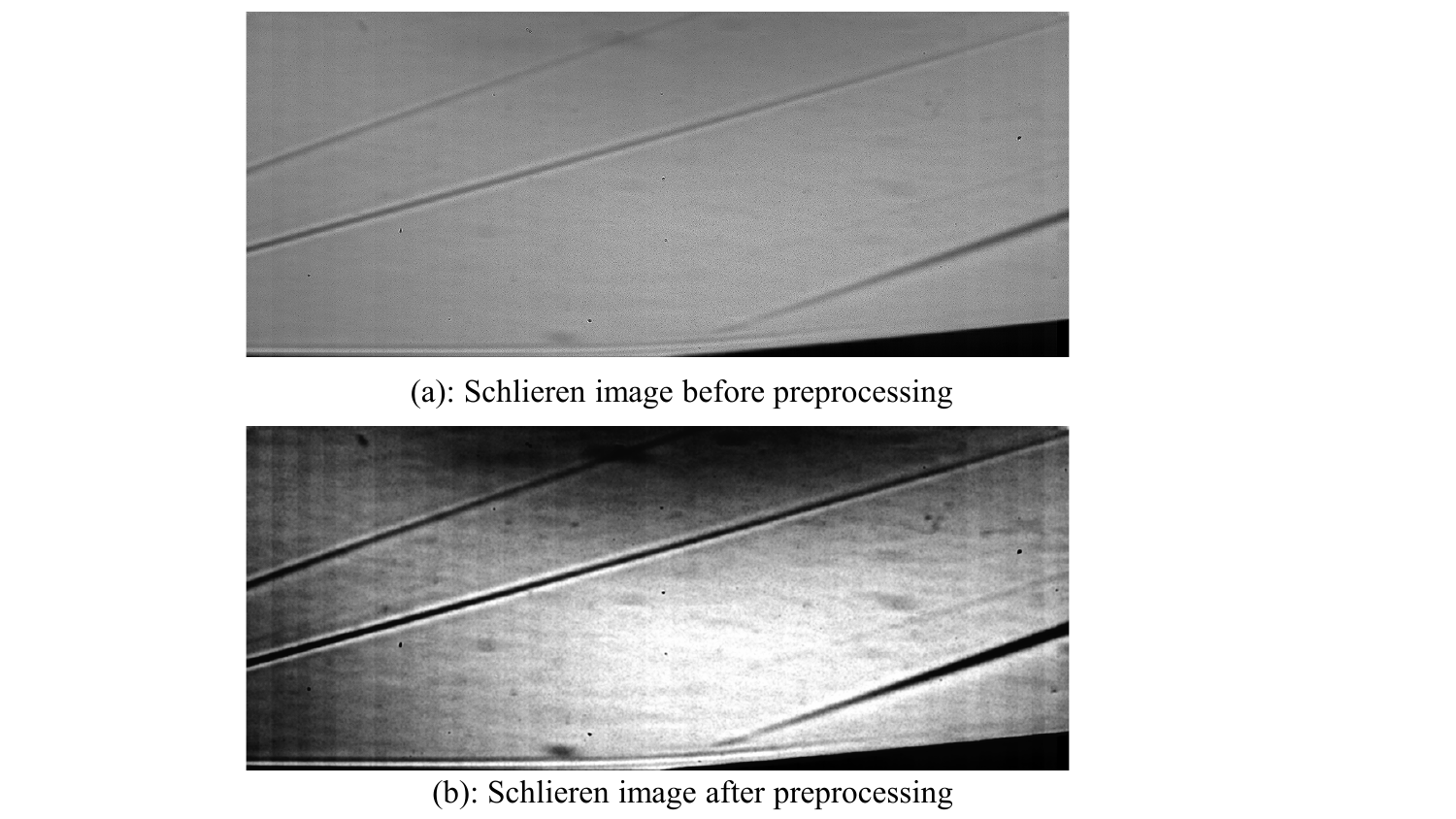}
	\caption{{The schlieren image before and after preprocessing for case 1, with $ Ma $ = 4,  $ Re   = 7.22\times 10^6$,  $ \alpha  $ = 6°, $ T_{\infty}$ = 70.7 K and  $ T_w /T_0 $ = 0.99.}}
	\label{figs:block-sliding-compare}
\end{figure}

 For each pixel location $(i,j)$ in the image, the gradient in the $y$-direction is computed as:

\begin{equation}
	  G_{y}(i,j) =  \sum_{m=-1}^{1} \sum_{n=-1}^{1}I(i+m,j+n)\times{K_{y}(m+1,n+1)}  
	\label{momentum-equation},
\end{equation}
where $I(i+m, j+n)$ is the grayscale value of the pixel at position $(i+m, j+n)$
 and the Sobel kernel $ K_y$ can be expressed by:
 
 \begin{equation}
	  K_y=\begin{bmatrix} -1  & -2 &-1\\ 0  & 0 & 0\\
      1  & 2 & 1
\end{bmatrix} 
	\label{momentum-equation}.
\end{equation}

Given that $ G_{y}$ = $\frac{\partial{I}} {\partial {y}}$ 
$ \propto$ $\frac{\partial{\rho}}{\partial \mathit{y}}$, we select  a representative pixel at $(i,j) = (64, 48)$ -- approximately the brightest pixel -- where the density gradient is taken as $\frac{\partial\rho}{\partial y} = 1.0~\text{kg/m}^4$. The density gradient in the $y$ direction across the entire schlieren image is then obtained by:
\begin{equation}
	  (\frac{\partial {\rho}}{\partial \mathit{y}})_{i,j}=(\frac{\partial I}{\partial \mathit{y}})_{i,j} \times \alpha'
	\label{momentum-equation},
\end{equation}
where the scaling factor $\alpha'$ is defined as:
\begin{equation}
	 \alpha'= \frac{\partial{\rho}} {\partial {y}} / \frac{\partial{I}} {\partial {y}} |_{(i,j)=(64,48)}
	\label{momentum-equation}.
\end{equation}
 The relative magnitude of the density field is subsequently reconstructed as:
\begin{equation}
\boldsymbol{\hat\rho} = \mathbf{T}  \cdot \frac{\partial \boldsymbol{\rho}}{\partial \mathit{y}} + \langle \mathbf{T}_{0}  \cdot \frac{\partial \boldsymbol{\rho}}{\partial \mathit{y}}\rangle_{x}  
	\label{momentum-equation},
\end{equation}
where $ \mathbf{T}  \cdot \frac{\partial \boldsymbol{\rho}}{\partial \mathit{y}} =\int_{0}^{y}\frac{\partial {\rho}}{\partial \mathit{y'}}dy'$ denotes the uncorrected density field containing the integration constant through trapezoidal numerical integration, $ \langle \mathbf{T}_{0}  \cdot \frac{\partial \boldsymbol{\rho}}{\partial \mathit{y}}\rangle_{x}=  \langle \rho(y_{0})\rangle_{x}$ represents the average density value of the reference row $y_{0}$=0. Finally, the relative  density ${\hat\rho}$ is normalized by its maximum and minimum values:
\begin{equation}
	  {\hat\rho}^{*}=\frac{({\hat\rho}-{\hat\rho}_{min})}{({\hat\rho}_{max} -{\hat\rho}_{min})} .
\end{equation}
Therefore, the normalized relative  density ${\hat\rho}^{*}$ is derived from the schlieren image.

\section{ Methodology }
\label{preliminary}

\subsection{ Neural network architectures }\label{subsec:splitting cell-linked list}
\begin{figure}[htb!]
	\centering
	\includegraphics[trim = 1cm 3.5cm 1cm 1cm, clip,width=0.99\textwidth]{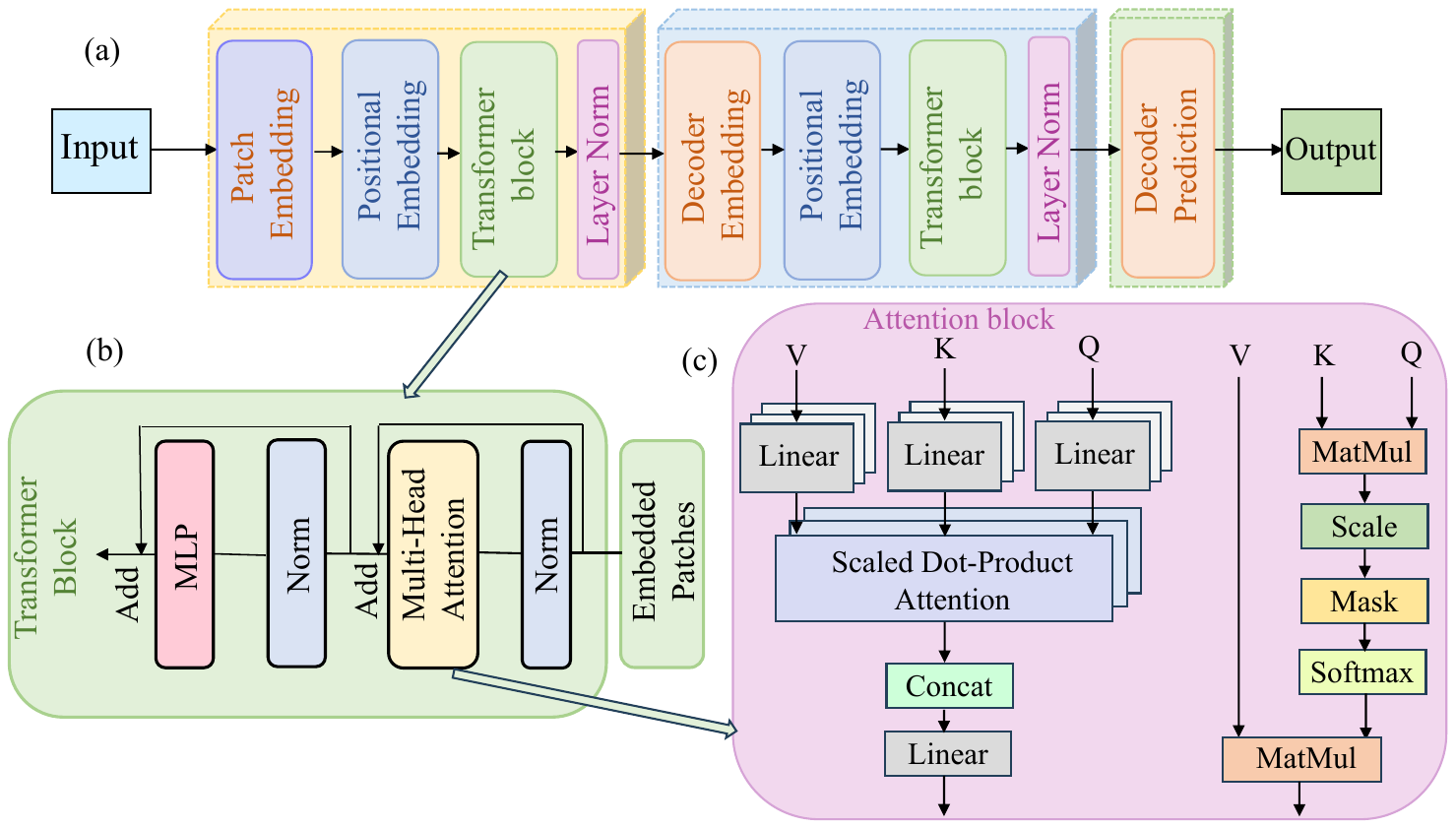}
	\caption{{The architecture of the Encoder-Decoder-based vision transformer network. (a): The network structure; (b): Details of the transformer block; (c):The attention block (multi-head attention) applied in the transformer block.}}
	\label{figs:block-sliding-compare}
\end{figure}

There are two neural network architectures applied for the prediction: an encoder-decoder-based vision transformer and a CNN-based attention network, as illustrated in Figures 6 and 7, respectively. 

Figure 6 presents the architecture of the encoder-decoder-based vision transformer. This model begins by dividing the input image $ \mathcal{X} \in  \mathbb{R}^{W \times  H \times C_{in}}$ into $N_{s} = W*H/P^{2}$ non-overlapping patches, where $W$ and $H$ denote the width and height of the image, $C_{in}$ denotes the number of input channels, and each patch is of size $P \times P$. To preserve spatial information, positional embeddings are incorporated into the patch sequence.  The resulting token sequence is processed by a transformer encoder to extract hierarchical features, followed by a transformer decoder that progressively upsamples the representation to reconstruct the output $\hat{ \textbf{y}} \in \mathbb{R}^{W \times  H \times C_{out}}$, where $C_{out}$ represents the output channels.
Both the transformer encoder and decoder blocks are composed of normalization (Norm) layes, an attention block (multi-head attention block), and a feed-forward network layer. The Norm layer  normalizes each token across its feature dimension, while the multi-head attention mechanism—the core component of the transformer—enables each patch in the sequence to attend to all other positions, thereby effectively capturing global context.

The  multi-head attention system operates through the following computational steps: First, the input is projected by three separate linear layers into the Query($Q$), Key($K$) and Value($V$) matrices. The output  for each $head_{i}$ is computed as:
\begin{equation}
	  head_{i}=Attention (Q_{i}, K_{i}, V_{i}) = softmax (\frac{Q_{i}K_{i}^{T}}{\sqrt{d_{K}}} ) V_{i}
	\label{momentum-equation},
\end{equation}
In this formulation: $Q_i K_i^T$  computes the attention score matrix (MatMul) and the result is scaled by $\frac{Q_i K_i^T}{\sqrt{d_k}}$ to stabilize gradients, where $d_k$ is the dimension of the query and key vectors; Optional mask matrix (Mask) may be applied in the decoder to ensure that the current position cannot observe future positions; The $\text{softmax}(\cdot)$ performs a softmax operation along each row (dim=-1) to convert scores into attention weights distributed probabilistically;
Then $\text{softmax}(\cdot) V_i$ performs a weighted sum of $V_i$ using the attention weights (the second MatMul). 
The outputs of all attention heads are subsequently concatenated along the feature dimension: 
\begin{equation}
	  MultiHead(Q,K,V) = Contact(head_{1}, head_{2},...head_{h})
	\label{momentum-equation}
\end{equation}
Finally, the concatenated result is projected through a linear layer $W^O$ to integrate information from all heads and restore the original model dimension.

\begin{figure}
	\centering
	\begin{subfigure}[b]{0.99\textwidth}
		\centering
		\includegraphics[trim = 0cm 0cm 0cm 0cm, clip,width=0.99\textwidth]{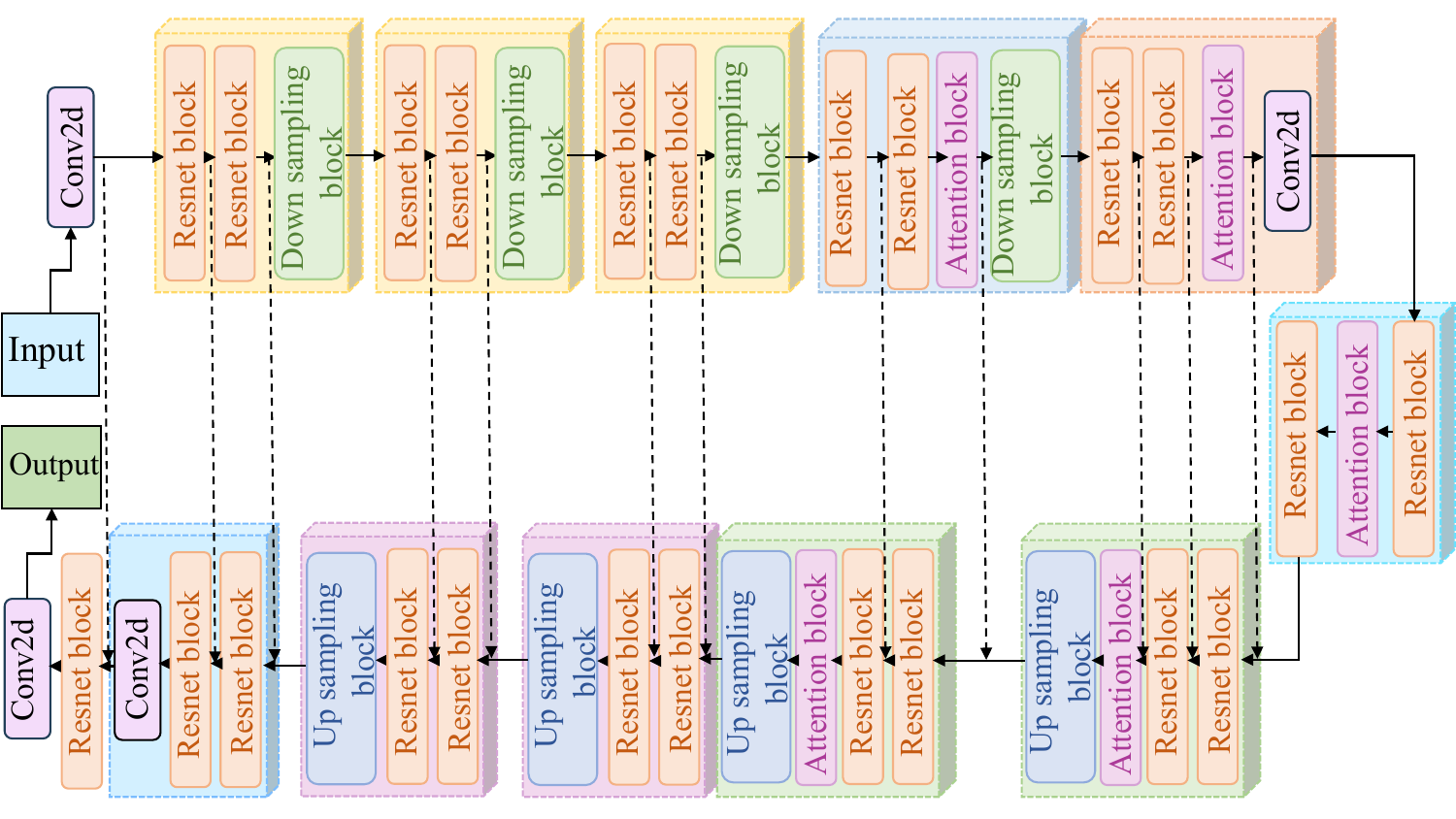}
		\caption{}
		\label{figs:re-constructed(a)}
	\end{subfigure}
        \begin{subfigure}[b]{0.99\textwidth}
		\centering
		\includegraphics[trim = 0cm 3cm 0cm 0cm, clip, width=\textwidth]{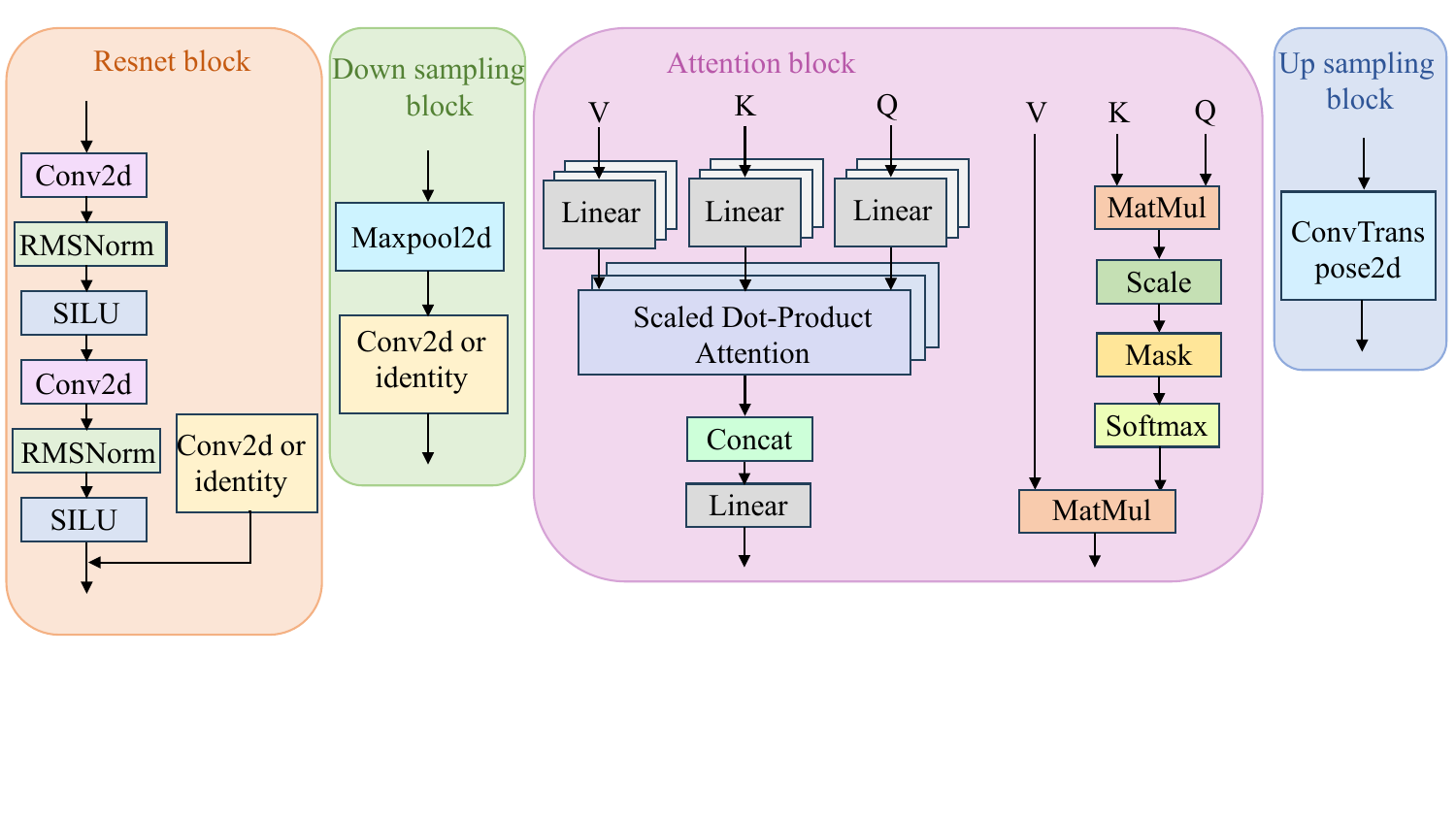}
		\caption{}
		\label{figs:re-constructed(a)}
	\end{subfigure}
	\caption{The architecture of the attention CNN network. (a): The network structure; Dashed line: skip connections; (b): Detail of each block. }
\end{figure}
Figure 7 illustrates the architecture of the attention-based CNN, detailing the components of each block. The model is composed primarily of Resnet blocks, attention blocks, down sampling blocks and up sampling blocks. The Resnet block is a modernized version built on the classic skip connection architecture. It comprises two parallel pathways: the main path and the shortcut path. The main path sequentially applies a Conv2d layer for feature extraction, an RMSNorm layer for normalization, and a SiLU activation function to introduce non-linearity; this sequence is then repeated. Meanwhile, the shortcut path employs a $1 \times 1$ Conv2d layer to project the input to match the output dimensions of the main path. If the input and output dimensions are already consistent, this path reduces to an identity connection, directly passing the input through. The final output is generated by summing the outputs from both paths.

The down-sampling and up-sampling processes  are connected with skip connections. These skip connections enable feature connection between the encoder and decoder, effectively preserving the feature information from the encoding layers to the corresponding decoding layers, thereby effectively preserving detailed information throughout the network. The attention block is implemented using a standard multi-head attention system, as previously introduced.

\begin{figure}[htb!]
	\centering
	\includegraphics[trim = 3cm 3.5cm 5.5cm 3cm, clip,width=0.89\textwidth]{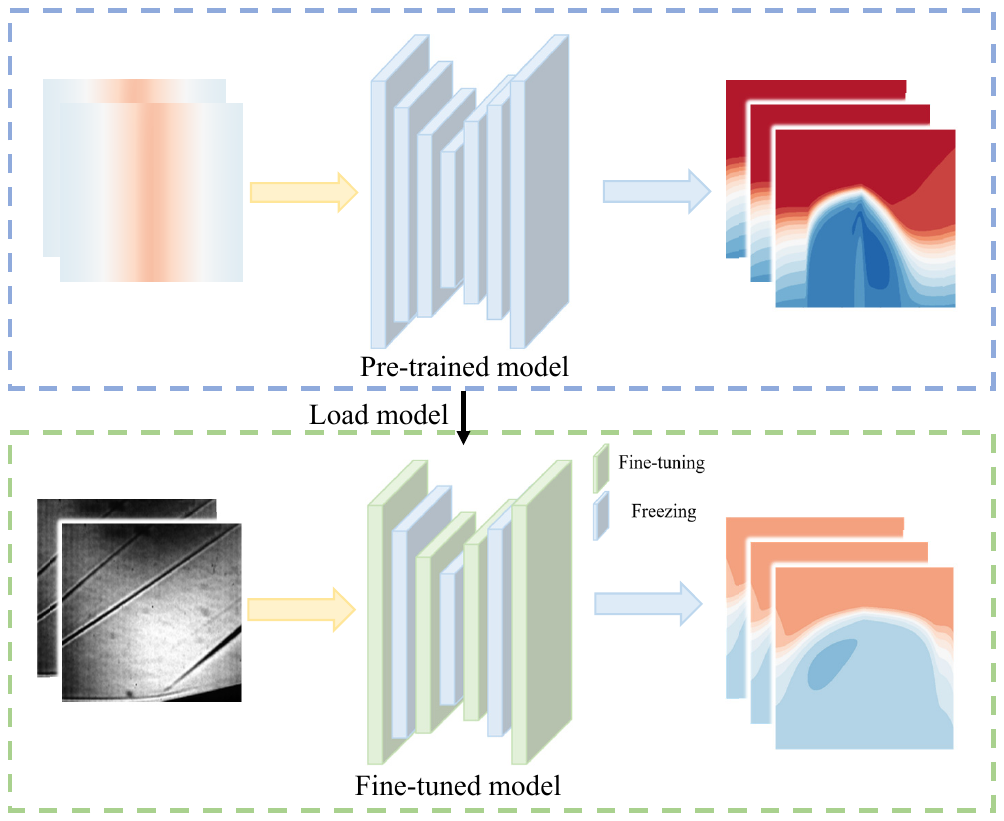}
	\caption{{The architecture of fine-tuned model acquisition process.}}
	\label{figs:block-sliding-compare}
\end{figure}
Figure 8 illustrates the procedure of getting the fine-tuned model, which mainly has two processes: Pre-training process (blue rectangular) and fine-tuning process (green rectangular). First, the neural network is pre-trained using the training dataset only contains CFD data. Then the pre-trained model is loaded and fine-tuned using a small set of schlieren images. In this stage, some layers of the network are frozen (parameters remain unchanged, shown in blue), while others are unfrozen and updated (shown in green) to adapt the model to the characteristics of the schlieren image data. The fine-tuned model is then used to generate the final output (velocity and pressure of the flow fields).


%

\subsection{ Training and evaluation }\label{subsec:splitting cell-linked list}
During training, the overall loss function $\mathcal{L}$ is defined as a weighted combination of a CFD-based loss $\mathcal{L}_{cfd}$ and an experiment-based loss $\mathcal{L}_{exp}$:
\begin{equation}
	  \mathcal{L} = \lambda\mathcal{L}_{cfd} + \mu \mathcal{L}_{exp} 
	\label{momentum-equation},
\end{equation}
where $\lambda$ and $\mu$ are the corresponding loss weights. The individual loss terms are given by:
\begin{equation}
	  \mathcal{L}_{cfd} = \frac{1}{N} ||\mathbf{y} - \mathbf{\hat{y}}||_2^2
	\label{momentum-equation},
\end{equation}
\begin{equation}
	  \mathcal{L}_{exp} = \frac{1}{N^{*}} ||\mathbf{y^{*}} - \mathbf{\hat{y}^{*}}||_2^2
	\label{momentum-equation},
\end{equation}
Here, $N$ and $N^{*}$ are the target amount of CFD and experiment, respectively; The vector $\mathbf{y}$ represents the ground truth CFD results, which include ($u$, $v$, $p$, $\rho$), while $\hat{\mathbf{y}}$ denotes the corresponding model predictions. Similarly, $\mathbf{y^{*}}$ refers to the experimental targets, comprising pressure readings from sensors and the relative density ${\hat\rho}^{*}$, with $\mathbf{\hat{y}^{*}}$ being the model predictions at the same sensor locations and density field. Thus, the density-related loss term actually is the difference in the relative density between the target and the prediction.


To validate the prediction accuracy of the proposed model, the error evaluation criteria are defined as follows:
\begin{equation}
	  \bm{\zeta}  = \frac{1}{N_{t}} \sum_{i=1}^{N_{t}} ||\mathbf{y}^{i}-\hat{\mathbf{y}}^{i}||_{1} 
	\label{momentum-equation},
\end{equation}
where $\bm{\zeta}$  represents the mean absolute error, $\mathbf{y}^{i}$ =$[u, v, p, \rho] $  and $\hat{\mathbf{y}}^{i}$ = $[\hat{u}, \hat{v}, \hat{p}, \hat{\rho}] \in $ $\hat{\mathbf{y}}^{i}$ represent the ground truth and predicted flow fields, respectively, for test cases  $i=1,2,...N_{t}$. During training, all physical quantities are normalized to the interval [0, 1] using min–max normalization for loss and error computation. For visualization purposes, the predicted flow fields are inversely transformed back to their original physical scales.
\begin{table}
\centering
\caption{Four models using different architectures or training datasets.}
\resizebox{0.95\textwidth}{!}{%
\tiny 
\begin{tabular}{cccc}
\hline
Model   &Architecture   &  training dataset &  training loss \\
\hline
A &ViT: encoder-decoder-based vision transformer & CFD & $\mathcal{L}_{cfd}$ \\
B &ViT: encoder-decoder-based vision transformer & CFD, experiment & $\mathcal{L}$   \\
C &Attention CNN: encoder-decoder-based attention network & CFD  & $\mathcal{L}_{cfd}$\\
D &Attention CNN: encoder-decoder-based attention network &  CFD, experiment & $\mathcal{L}$ \\

\hline
\end{tabular}%
}
\end{table}
\begin{table}
\centering
\caption{The process of fine-tuning model generation.}
\resizebox{0.55\textwidth}{!}{%
\tiny 
\begin{tabular}{ccc}
\hline
Model     &   dataset &  training loss \\
\hline
Pre-trained  & CFD & $\mathcal{L}_{cfd}$ \\
Fine-tuned  & Schlieren image & $\mathcal{L}_{exp}$ (only density part)   \\

\hline
\end{tabular}%
}
\end{table}

Table 2 presents the training models based on different architectures, which are trained using either CFD-only or hybrid CFD–experimental datasets. Table 3 summarizes the training losses employed during both the pre-training and fine-tuning stages. 

The training process is conducted on an NVIDIA GeForce RTX 4090 GPU. The Adam optimizer \cite{kingma2014adam} is used with an initial learning rate of 1e-4. During fine-tuning, the learning rate is set to 1e-5.

\section{ Results and analysis }
\subsection{Comparative performance evaluation of the models}
The non-uniform physical field is first transformed into a uniform computational space  through coordinate transformation. Subsequently, various network architectures are evaluated using different training datasets.

As summarized in Table 4, which lists the prediction errors for all ten test cases across models A to D with the lowest error marked in blue, the inclusion of experimental data leads to a slight reduction in overall prediction error.  This improvement is particularly noticeable for density and pressure, as the Kulite pressure sensors and schlieren images provide supplementary information on surface pressure and relative density. For example, the pressure error decreases from 0.013520 to 0.011283 (a 16.5\% reduction) when comparing model A to model B, and from 0.012549 to 0.010962 (a 12.6\% reduction) from model C to model D. Similarly, the density error declines from 0.004458 to 0.004132 (model A to B) and from 0.004607 to 0.004266 (model C to D), corresponding to reductions of 7.3\% and 7.4\%, respectively. These results indicate that incorporating experimental data helps achieve higher accuracy in flow field reconstruction. Furthermore, a separate comparison between models A and C, or between models B and D, reveals that the attention CNN architecture adopted in this study achieves higher overall accuracy than the ViT.

\begin{table}
\centering
\caption{The average errors of  models A-D on different training datasets for all test cases.}
\resizebox{0.9\textwidth}{!}{%
\tiny 
\begin{tabular}{cccccc}
\hline
Models   & $\bm{\zeta}$              &$\bm{\zeta}_{u}$            & $\bm{\zeta}_{v}$    &  $\bm{\zeta}_{p}$   &  $\bm{\zeta}_{\rho}$  \\
\hline
 A & 0.014654 & 0.032091 & 0.086692 & 0.013520 & 0.004458 \\
 B & 0.013968 & 0.032573 &  {\color{blue}0.007886} & 0.011283 & {\color{blue}0.004132} \\
 C & 0.013989 & 0.030505 & 0.008294 & 0.012549 & 0.004607 \\
 D & {\color{blue}0.013109} & {\color{blue}0.029218} &  0.007990 & {\color{blue}0.010962} & 0.004266 \\
\hline
\end{tabular}%
}
\end{table}
Then the prediction results of the flow details with models A-D are demonstrated using the case of $ Ma $ = 6.36,  $ Re   = 4.26\times 10^6$,  $ \alpha  $ = 14.2°, $ T_{\infty}$ = 127 K and  $ T_w /T_0 $ = 0.09. Figures 9 to 12 present a comparative visualization of the velocity, pressure, and density fields between the ground truth and the predictions of models A–D.  Generally, the integration of experimental data enables a more accurate capture of the boundary layer near the corner, as shown in Figures 9(k)–9(n), 10(k)–10(n), 11(k)–11(n), and 12(k)–12(n). Also, the proposed attention CNN achieves a lower prediction error than the ViT. The reasons for this performance advantage can be attributed to the following aspects:

First, regarding the characteristics of flow field data, although hypersonic flow fields—especially near the compression corner—exhibit complex flow features, variations in physical quantities such as pressure and density gradients remain localized and smooth across most regions. In addition, distinct structures such as shock waves and shear layers also display clear local features. The inherent advantage of our approach lies in its hybrid architecture. While the convolutional backbone excels at capturing these localized and smooth physical patterns, the incorporated attention mechanism strategically augments this capability by enabling the model to focus on critical global structures and long-range interactions, which are also evident in complex phenomena such as shock reflections. This design establishes a synergistic balance between local precision and global contextual awareness. In contrast, although the ViT employs a powerful global self-attention mechanism, its patch-based processing strategy inherently disrupts fine-grained spatial continuity—a property essential for accurate reconstruction of physical fields.

\begin{figure}[htb!]
	\centering
	\includegraphics[trim = 3.7cm 3.5cm 4.2cm 3.5cm, clip,width=\textwidth]{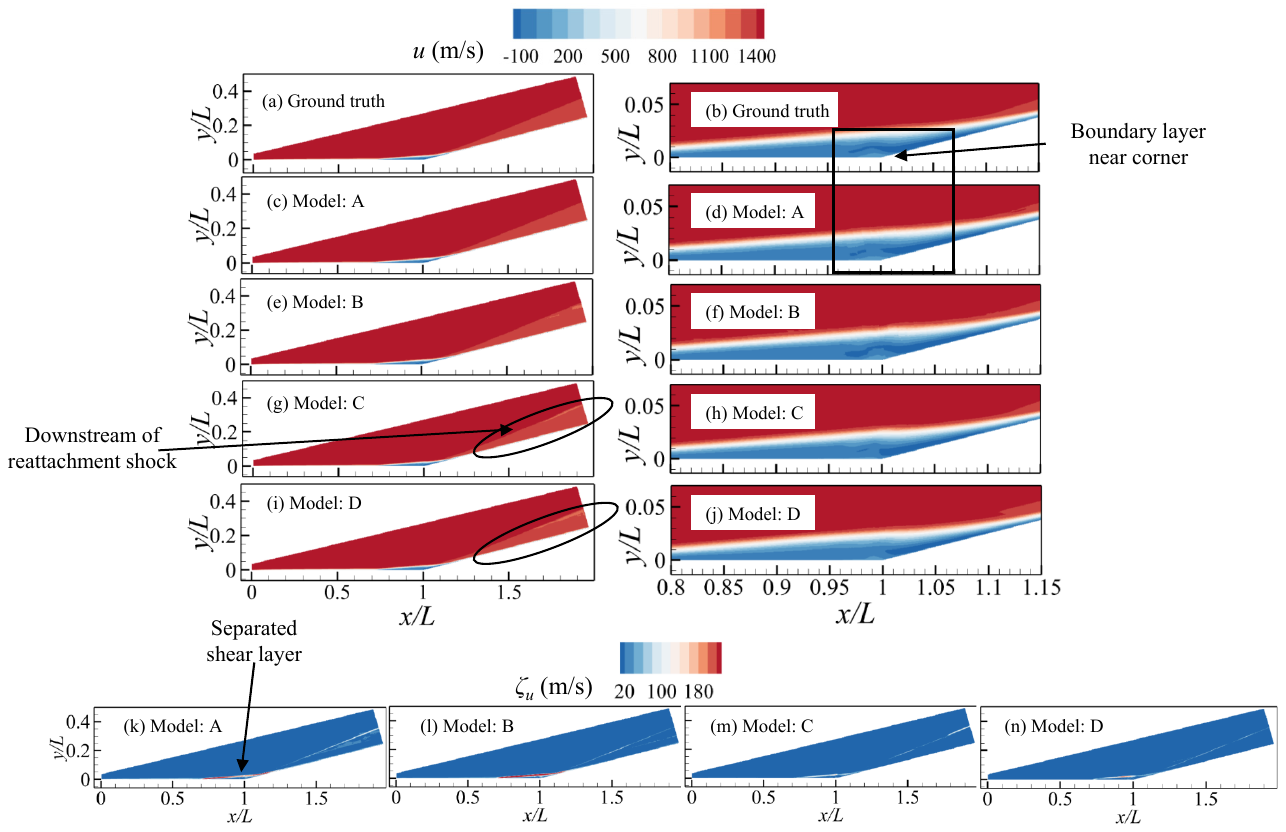}
	\caption{{Comparison of $u$ velocity fields of ground truth and models A-D at $Ma=6.36$, $Re=4.26\times10^6$, $\alpha=14.2^\circ$, $ T_{\infty}$ = 127 K, and  $ T_w /T_0 $ = 0.09.  (a)-(b) Ground truth; (c)-(j) Models A-D; (k)-(n) Prediction errors.}}
	\label{figs:block-sliding-compare}
\end{figure}

Figures 10(c) and 10(e) demonstrate that incorporating experimental data enhances the prediction accuracy in the downstream flow field and mitigates patch-induced discontinuities. This improvement is further confirmed by comparing Figures 10(g) and 10(i), which illustrate the benefits of incorporating experimental data for the downstream region. In the enlarged views, model B (Figure 10(f)) captures the separation bubble more accurately than model A (Figure 10(d)), demonstrating the value of experimental data in refining flow field predictions. Once again, Figures 10(k)–10(n) indicate that the inclusion of experimental data reduces prediction errors near the regions downstream of the reattachment shock and the separation shock. Moreover, under identical training conditions, the proposed attention CNN achieves lower errors than the ViT used in this study.
\begin{figure}[htb!]
	\centering
	\includegraphics[trim = 3.7cm 3.2cm 4.2cm 3.5cm, clip,width=\textwidth]{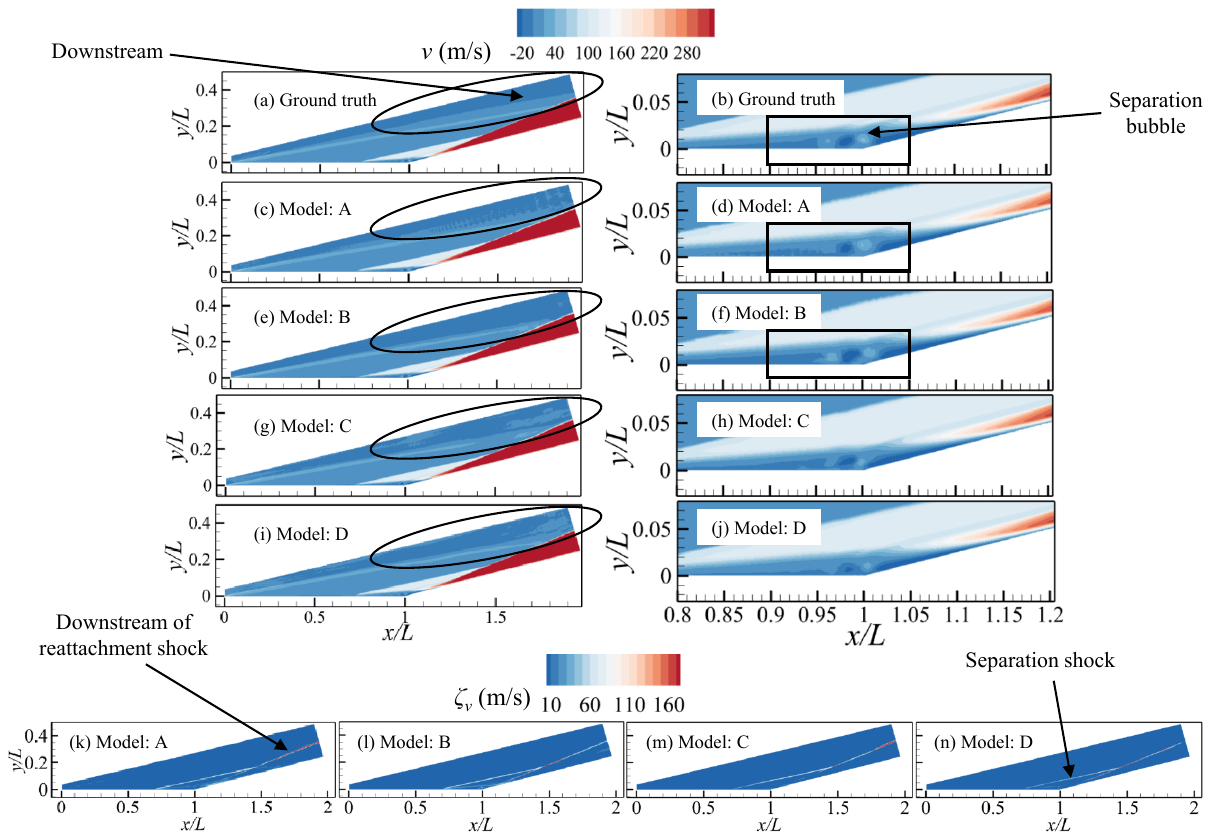}
	\caption{{Comparison of $v$ velocity fields of ground truth and models A-D at $Ma=6.36$, $Re=4.26\times10^6$, $\alpha=14.2^\circ$, $ T_{\infty}$ = 127 K, and  $ T_w /T_0 $ = 0.09.  (a)-(b) Ground truth; (c)-(j) Models A-D; (k)-(n) Prediction errors.}}
	\label{figs:block-sliding-compare}
\end{figure}

Figures 11(c)–11(f) reveal that incorporating experimental wall pressure measurements and schlieren data improves predictive accuracy in the near-wall expansion wave and reflected expansion wave regions. For the attention CNN architecture (Figures 11(g)–11(j)), the integration of experimental data brings only a marginal enhancement, with the most noticeable improvement occurring downstream of the reattachment shock, where the prediction becomes smoother and aligns more closely with the ground truth. Similar to the $v$ velocity fields, the use of experimental data reduces prediction errors near the separation shock and the region downstream of the reattachment shock. 
\begin{figure}[htb!]
	\centering
	\includegraphics[trim = 3.7cm 3.2cm 4.2cm 3.5cm, clip,width=\textwidth]{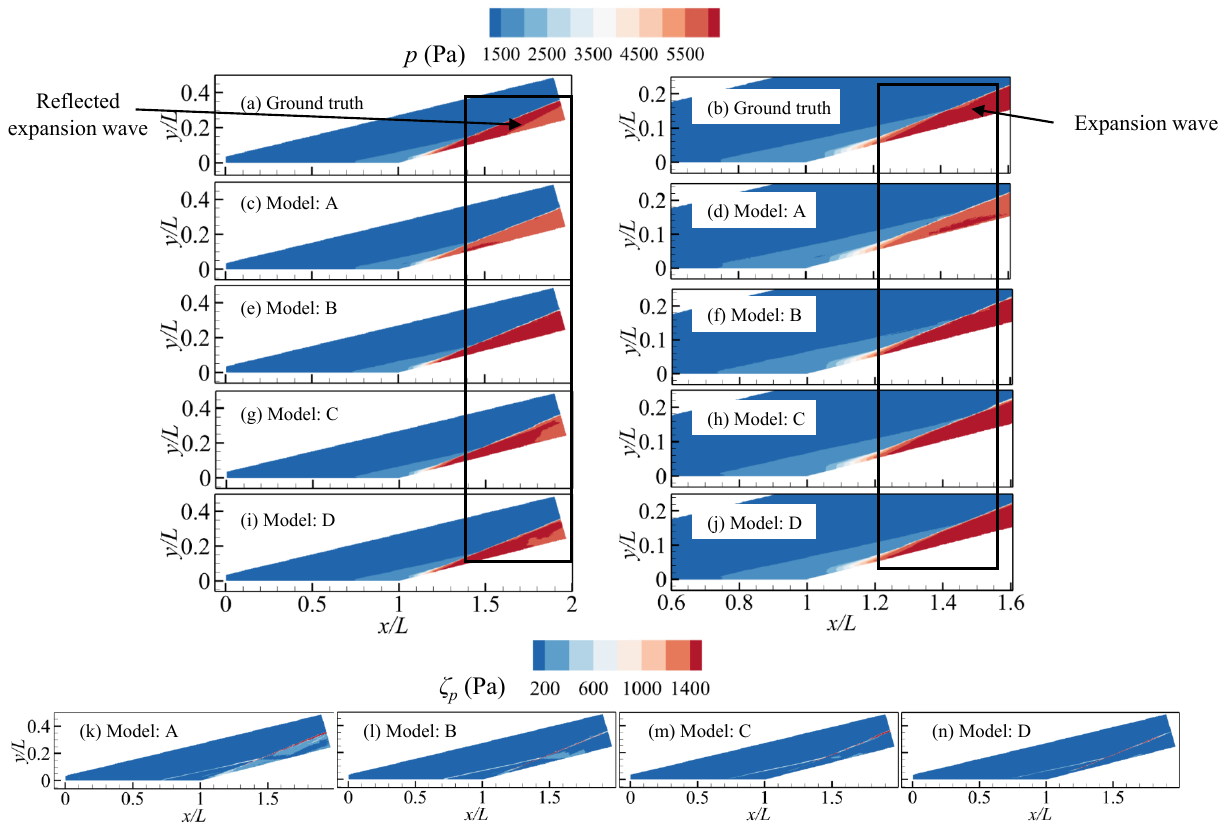}
	\caption{{Comparison of pressure fields of ground truth and models A-D at $Ma=6.36$, $Re=4.26\times10^6$, $\alpha=14.2^\circ$, $ T_{\infty}$ = 127 K, and  $ T_w /T_0 $ = 0.09.  (a)-(b) Ground truth; (c)-(j) Models A-D; (k)-(n) Prediction errors.}}
	\label{figs:block-sliding-compare}
\end{figure}

Figures 12(d) and 12(f) show that introducing schlieren data alleviates patch-induced discontinuities in the density field near the corner. Meanwhile, Figures 12(g) and 12(i) demonstrate that the density field downstream of the reattachment shock appears smoother when schlieren information is incorporated. Consequently, as evidenced in Figures 12(k)–12(n), fusing schlieren data leads to a lower overall prediction error in the density field.
\begin{figure}[htb!]
	\centering
	\includegraphics[trim = 3.7cm 3.2cm 4.2cm 3.5cm, clip,width=\textwidth]{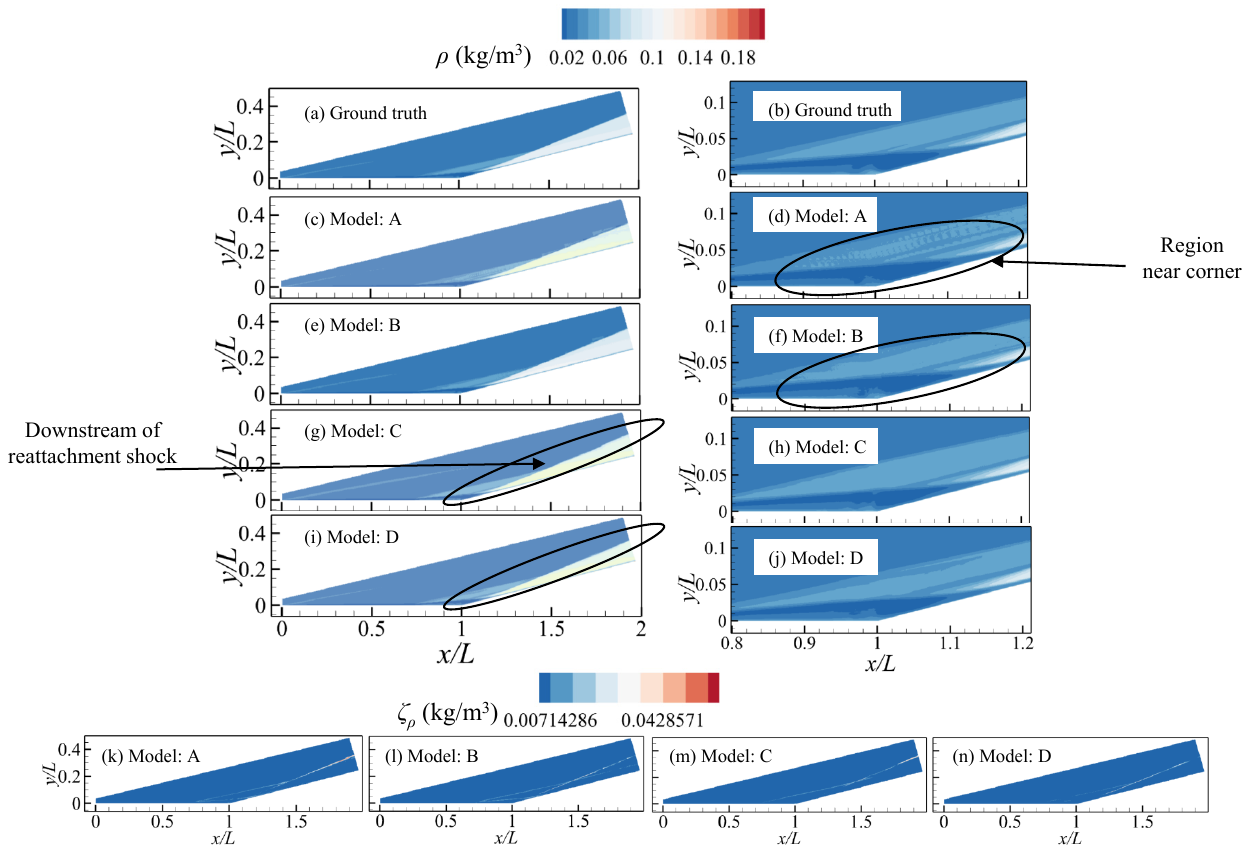}
	\caption{{Comparison of density fields of ground truth and models A-D at $Ma=6.36$, $Re=4.26\times10^6$, $\alpha=14.2^\circ$, $ T_{\infty}$ = 127 K, and  $ T_w /T_0 $ = 0.09.  (a)-(b) Ground truth; (c)-(j) Models A-D; (k)-(n) Prediction errors.}}
	\label{figs:block-sliding-compare}
\end{figure}

Figure 13 compares the density gradient magnitude of the ground truth with that of models A-D under the flow conditions of $Ma=6.36$, $Re=4.26\times 10^6$, $\alpha=14.2^\circ$, $T_{\infty}=127$ K, and $T_w/T_0=0.09$. The contours show that both models C and D accurately capture the leading edge of the separation shock. In the interaction region between the separation and reattachment shocks, model D exhibits slightly better performance than model C, attributable to the integration of the experimental schlieren images.

\begin{figure}[htb!]
	\centering
	\includegraphics[trim = 3cm 3.7cm 6cm 3cm, clip,width=0.99\textwidth]{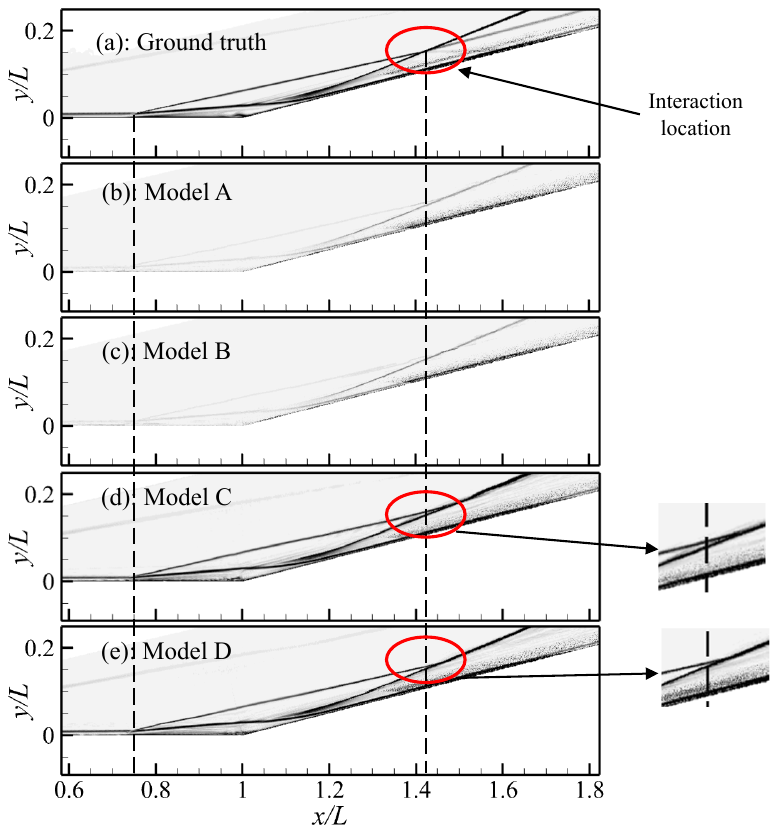}
	\caption{{Contours of density gradient magnitudes of ground truth and models A-D, at $ Ma $ = 6.36,  $ Re   = 4.26\times 10^6$,  $ \alpha  $ = 14.2°, $ T_{\infty}$ = 127 K and  $ T_w /T_0 $ = 0.09.}}
	\label{figs:block-sliding-compare}
\end{figure}
Overall, model D, which is based on the attention CNN architecture and incorporates experimental data, achieves superior predictive accuracy and captures fine-grained flow details more effectively. Consequently, it will be employed for the subsequent analysis of the flow physics.

\subsection{Physical analysis based on model D}
The transition of a two-dimensional separated flow to a three-dimensional state, driven by intrinsic global instability, can be quantitatively characterized through global stability analysis \cite{hao2021occurrence,theofilis2000origins, robinet2007bifurcations,theofilis2011global,hildebrand2018simulation}. To this end, Figure 14 displays the GSA results for the least stable mode from DNS and model D, showing strong agreement and confirming a globally stable flow at $Ma=6.61$, $Re=4.92\times 10^6$, $\alpha=11.1^\circ$, $T_{\infty}=113$ K, and $T_w/T_0=0.80$. Figure 15 presents the global stability results for the globally unstable flow field with secondary separation. From previous research, we know that the global instability occurs right before secondary separation \cite{hao2021occurrence}. By establishing global stability, this result strategically guides the subsequent mechanism analysis. It enables the exclusion of global instability as a source of three-dimensionality, thus directing focus toward the effects of convective instability on the flow field.

\begin{figure}
	\centering
	\begin{subfigure}[b]{0.49\textwidth}
		\centering
		\includegraphics[trim = 3cm 1cm 4cm 1cm, clip, width=\textwidth]{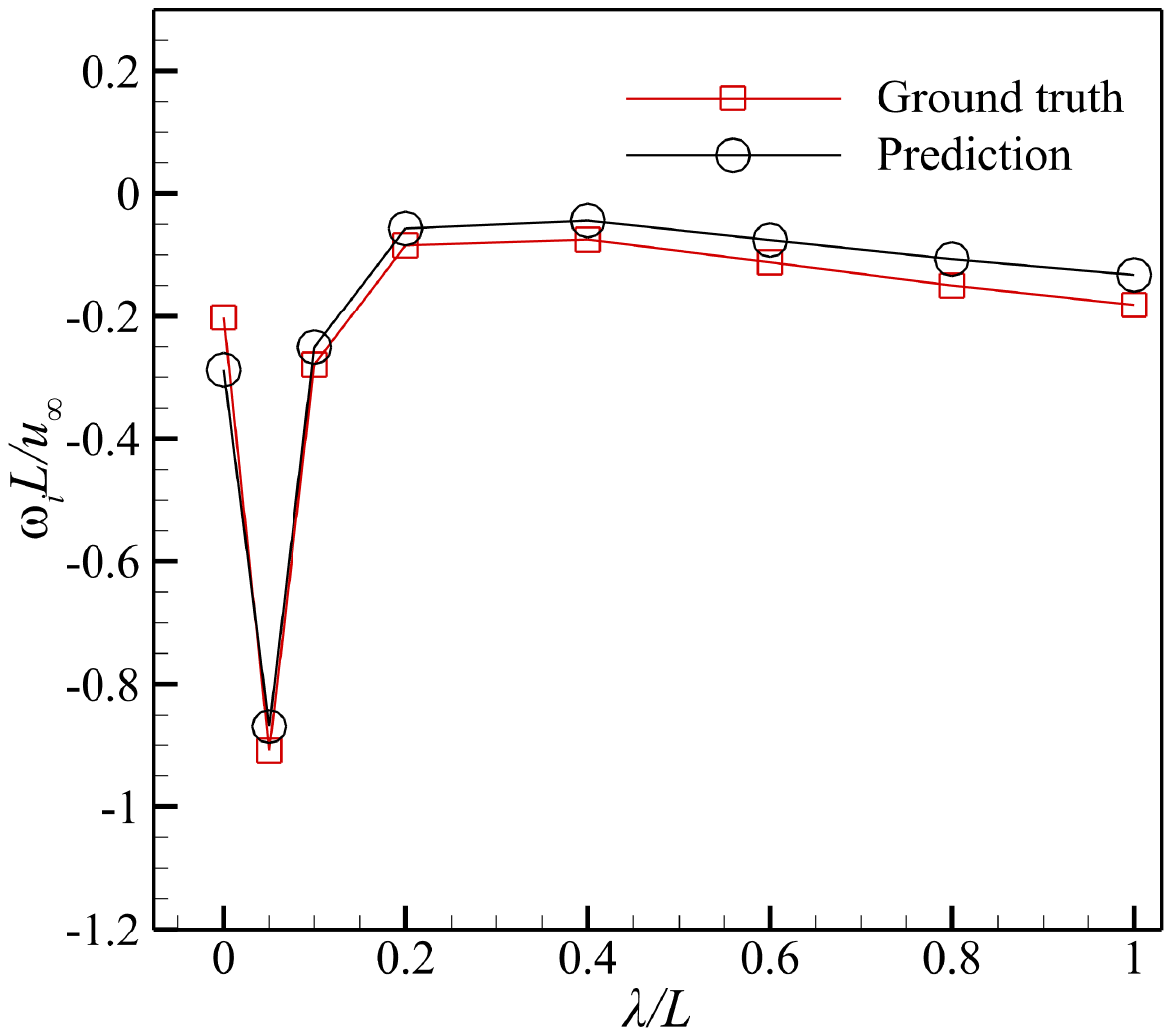}
		\caption{Growth rates of the most unstable mode}
		\label{figs:re-constructed(a)}
	\end{subfigure}
        \begin{subfigure}[b]{0.49\textwidth}
		\centering
		\includegraphics[trim = 3cm 1cm 4cm 1cm, clip, width=\textwidth]{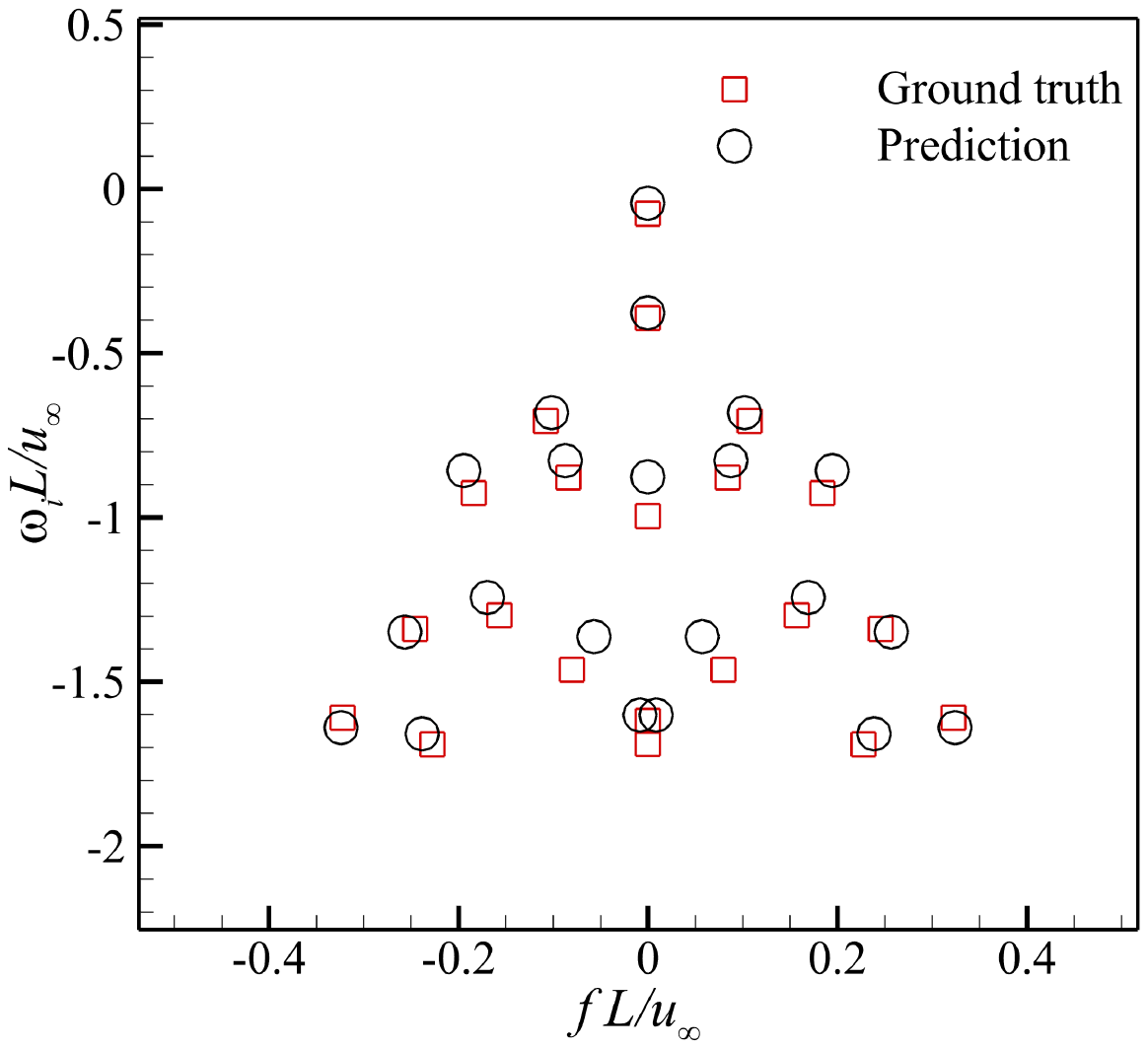}
		\caption{Eigenvalue spectra at $\lambda/L$ = 0.4}
		\label{figs:re-constructed(a)}
	\end{subfigure}

	\caption{Global instability analysis for the case at $ Ma $ = 6.61,  $ Re   = 4.92\times 10^6$,  $ \alpha  $ = 11.1°, $ T_{\infty}$ = 113 K and  $ T_w /T_0 $ = 0.80.  
		(a):  Growth rates of the most unstable mode as a function of spanwise wavelength; 
		(b): Eigenvalue spectra at $\lambda/L$ = 0.4.}
\end{figure}

\begin{figure}[htb!]
	\centering
	\includegraphics[trim = 3cm 1cm 3cm 2.5cm, clip,width=0.55\textwidth]{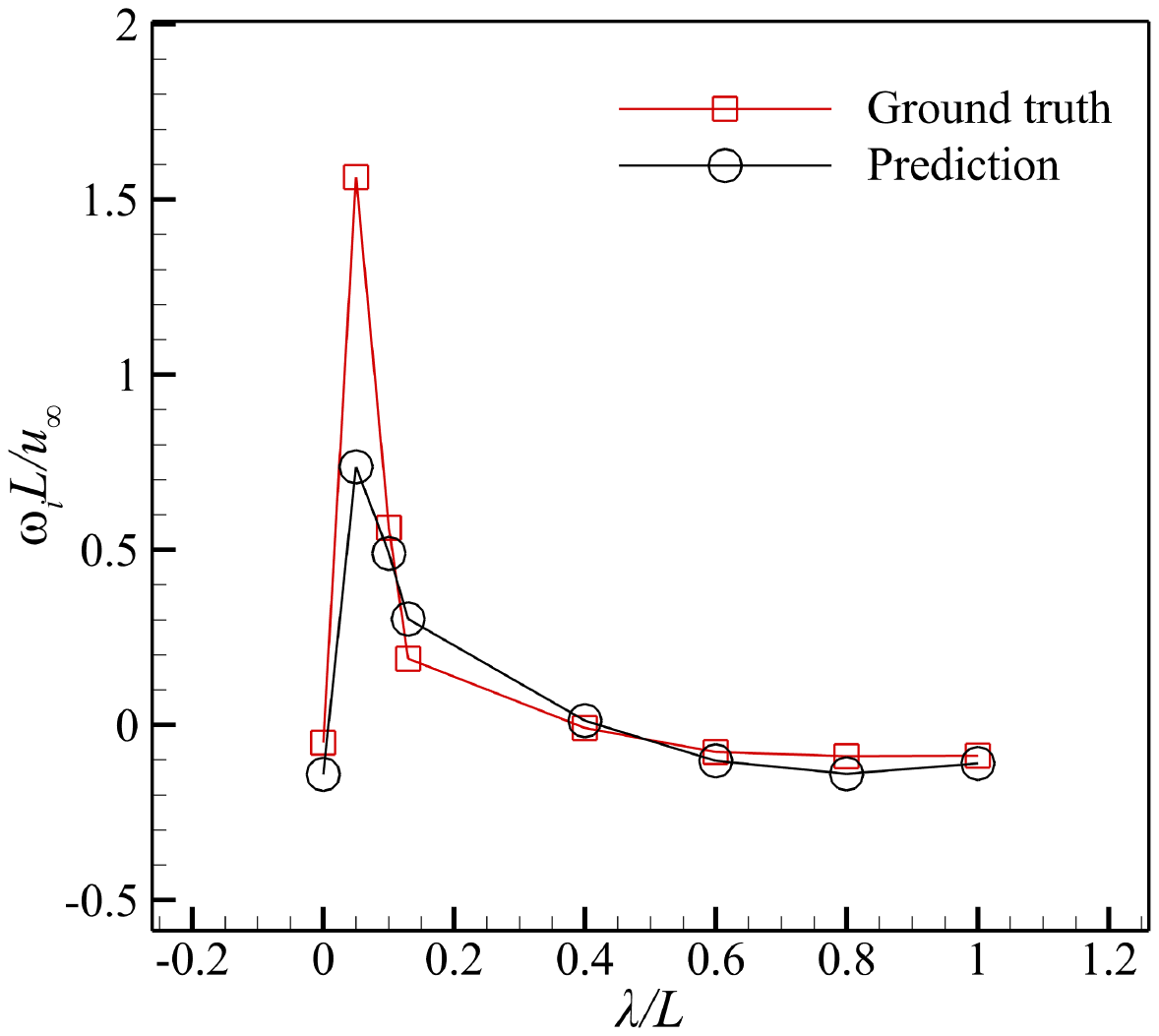}
	\caption{Growth rates of the most unstable mode as a function of spanwise wavelength for the case at $ Ma $ = 6.36,  $ Re   = 4.26\times 10^6$,  $ \alpha  $ = 14.2°, $ T_{\infty}$ = 127 K and  $ T_w /T_0 $ = 0.09.}
	\label{figs:block-sliding-compare}
\end{figure}

\begin{figure}
	\centering
         \begin{subfigure}[b]{1\textwidth}
		\centering
		\includegraphics[trim = 2.8cm 13.3cm 3cm 0cm, clip, width=0.6\textwidth]{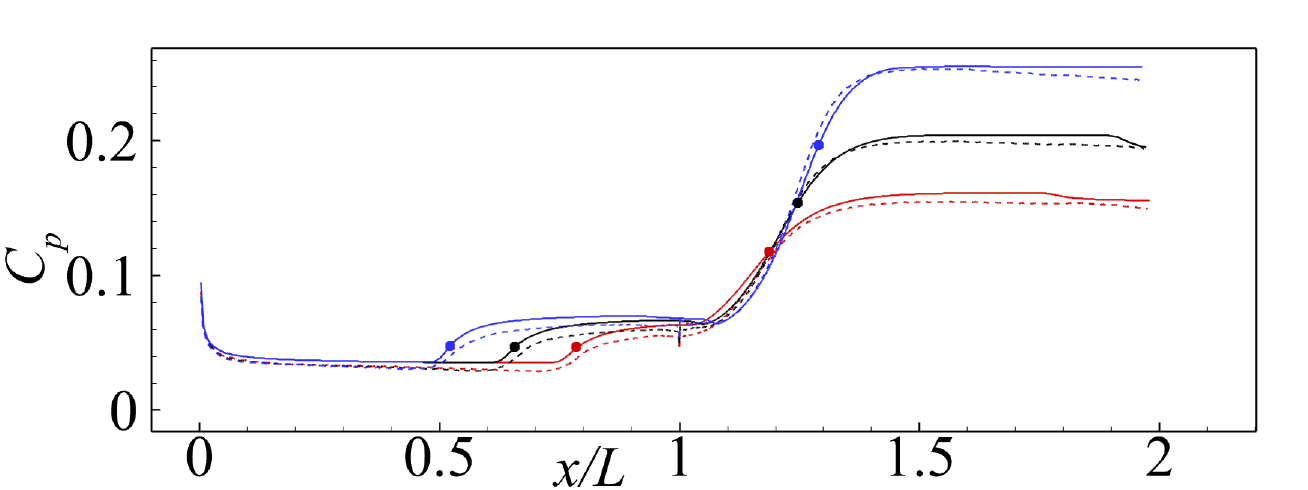}
        \caption{$ T_w /T_0 $ = 0.27}
        \end{subfigure}
                 \begin{subfigure}[b]{1\textwidth}
		\centering
		\includegraphics[trim = 2.8cm 13.3cm 3cm 0cm, clip, width=0.6\textwidth]{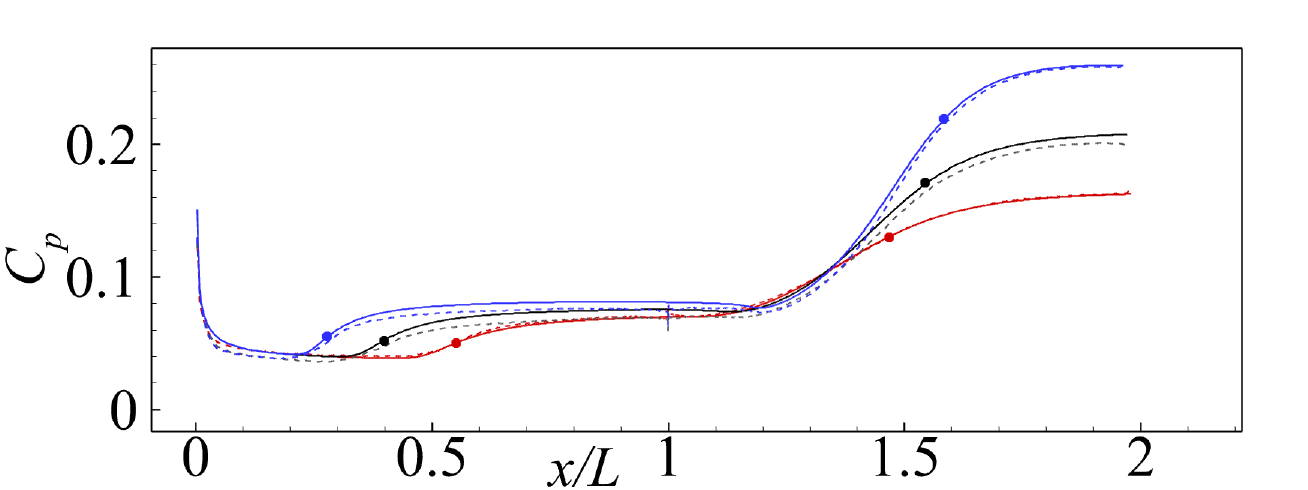}
        \caption{$ T_w /T_0 $ = 0.80}
        \end{subfigure}
                 \begin{subfigure}[b]{1\textwidth}
		\centering
		\includegraphics[trim = 2.8cm 13.3cm 3cm 0cm, clip, width=0.6\textwidth]{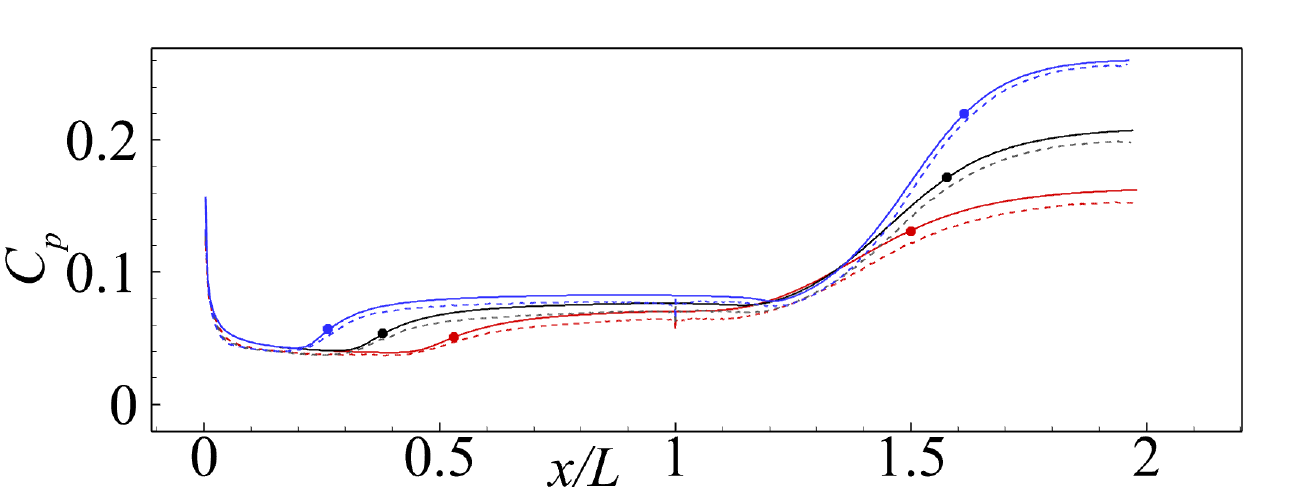}
        \caption{$ T_w /T_0 $ = 0.86}
        \end{subfigure}
	\caption{Distributions of the surface pressure coefficient from the ground truth and model D at $ Ma $ = 6.61 and  $ Re  $ = $4.92\times 10^6$, with various ramp angles and wall temperature ratios. Red, $ \alpha  $ = 11.1°; black, $ \alpha  $ = 13.1°; blue, $ \alpha  $ = 15.1°. Solid lines: ground truth; dashed lines: predictions. Closed circles denote separation and attachment points.}
\end{figure}
Figure 16 presents the wall pressure coefficient from the ground truth and model D under different wall temperatures (sub-figures) and ramp angles (line colors) at $Ma=6.61$, $Re=4.92\times 10^6$,  $T_{\infty}=113$ K. The wall temperature ratio $ T_w /T_0 $ = 0.27 corresponds to a wall temperature of $ T_w $ = 293 K, while the ratio of $ T_w /T_0 $ = 0.86 denotes the adiabatic wall condition, determined by a recovery factor of the Prandtl number $ P_r^{0.5} $. The surface pressure distribution exhibits a characteristic profile: An initial rise upstream of separation governed by free interaction \cite{chapman1958investigation}, followed by a plateau which increases with $ \alpha $ and $ T_w /T_0 $. A subsequent rise near reattachment leads to a peak value consistent with oblique shock theory. For the three cases in Figure 16(a), the reattachment shock interacts the separation shock in the downstream domain, inducing the final downstream drops of pressure induced by expansion waves. The results show that the predicted $C_p$ matches well with the ground truth and capture the characteristic \textquotedblleft dip\textquotedblright\, between the surface pressure plateau and peak, which indicates the presence of a pressure gradient in the corner region \cite{hao2021occurrence}. The \textquotedblleft dip\textquotedblright\,  becomes stronger with the increase of $\alpha$ so that the reverse flow boundary layer cannot resist the adverse pressure gradient and forms the secondary separation \cite{hao2021occurrence}.
\begin{figure}[htb!]
	\centering
	\includegraphics[trim = 1cm 6cm 2cm 7.5cm, clip,width=0.65\textwidth]{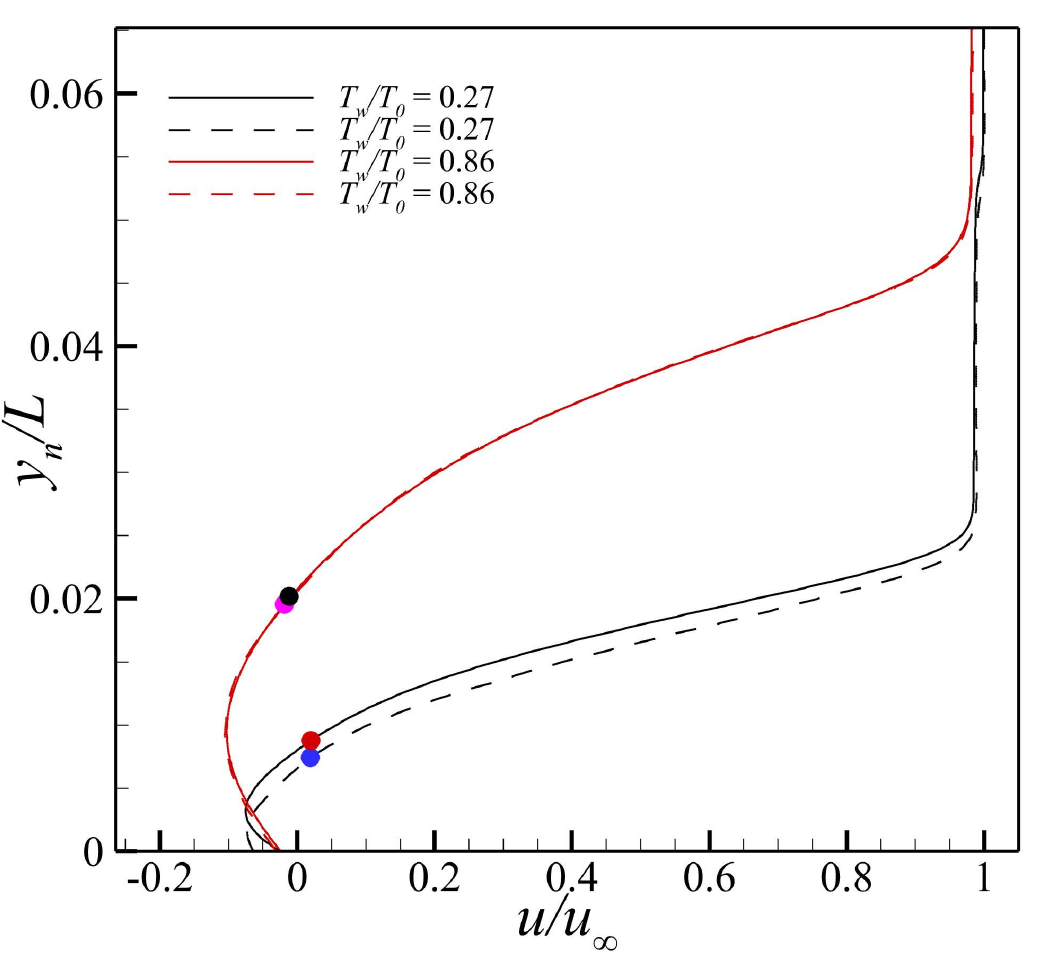}
	\caption{Streamwise velocity distributions along wall normal direction through the separation vortex cores. Solid lines: ground truth; Dashed lines: prediction results. Closed circles: core of vortex.}
	\label{figs:block-sliding-compare}
\end{figure}

As shown in Figure 17, the streamwise velocity profiles along the wall-normal direction through the separation vortex cores at $ Ma $ = 6.61,  $ Re   = 4.92\times 10^6$,  $ \alpha  $ = 11.1° are compared between the ground truth and model D. The vortex cores are marked by closed circles, with the reverse flow region located beneath it. The results demonstrate that model D accurately captures both the streamwise velocity distributions and the positions of the vortex core. The sharp velocity gradients above the vortex cores correspond to the shear layers originating from the flow separation. Furthermore, an increase in wall temperature is found to enhance the maximum reverse flow velocity in the region between the vortex core and the wall \cite{hao2021occurrence}.
\begin{figure}
	\centering
         \begin{subfigure}[b]{1\textwidth}
		\centering
		\includegraphics[trim = 4.5cm 3cm 6cm 13cm, clip, width=0.9\textwidth]{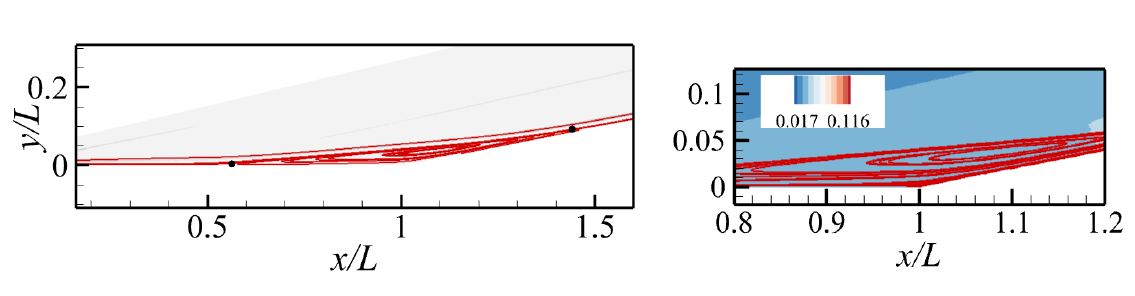}
        \caption{$ \alpha $ = 11.1°}
        \end{subfigure}
                 \begin{subfigure}[b]{1\textwidth}
		\centering
		\includegraphics[trim = 4.5cm 3cm 6.3cm 13cm, clip, width=0.9\textwidth]{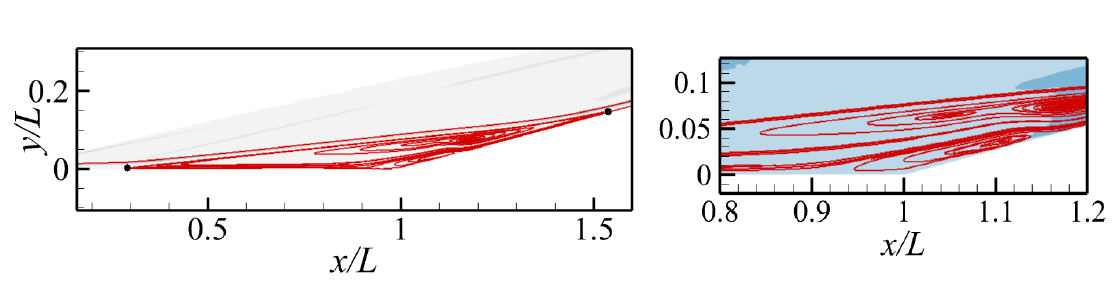}
        \caption{$ \alpha $ = 15.1°}
        \end{subfigure}
                 \begin{subfigure}[b]{1\textwidth}
		\centering
		\includegraphics[trim = 4cm 3cm 6.3cm 13cm, clip, width=0.9\textwidth]{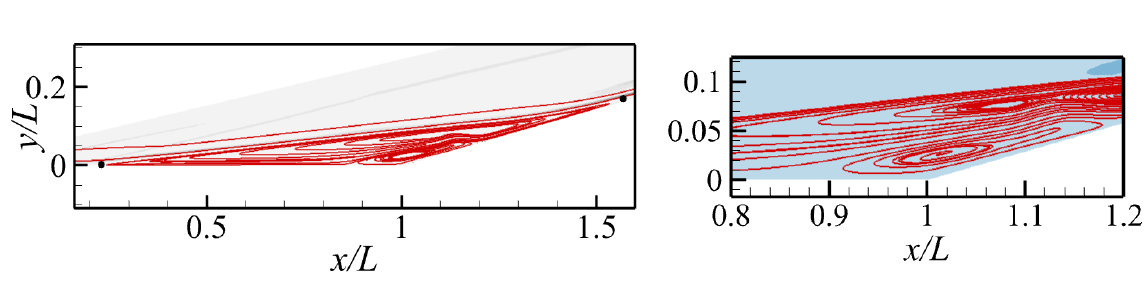}
        \caption{$ \alpha $ = 16.1°}
        \end{subfigure}
	\caption{Flow fields at $ Ma $ = 6.61,  $ Re  $ = $4.92\times 10^6$, and $ T_w /T_0 $ = 0.80: contours of density gradient magnitude (left) and non-dimensional pressure (right) with streamlines superimposed for (a)$ \alpha  $ = 11.1°, (b)$ \alpha  $ = 15.1°, (c)$ \alpha  $ = 16.1°.  Closed circles in the left figures indicate separation and reattachment positions.}
\end{figure}

Figure 18 presents contours of the density gradient magnitude and pressure (normalized by ${\rho }_{\infty }{u}_{\infty }^{2}$), along with streamlines, for flow conditions of $Ma=6.61$, $Re=4.92\times 10^6$, $T_{\infty}=113$ K, and $T_w/T_0=0.80$. The evolution of the flow field with increasing ramp angle is characterized as follows: At $ \alpha  $ = 11.1°, no secondary separation is observed, and the vortex core is located above the ramp surface downstream of the corner. As $ \alpha  $ increases to 15.1°, the primary separation bubble splits into two distinct vortices. With a further increase in the ramp angle to 16.1°,  the secondary separation bubble expands significantly in both the streamwise and wall-normal directions.

\subsection{Inference performance of fine-tuned model }
Based on Sections 4.1 and 4.2, models with either the attention CNN or ViT architecture are employed for fine-tuning.  The process begins with a model pre-trained exclusively on CFD data—identical to model A or model C. This pre-trained model is then fine-tuned using schlieren images. Here we list case 9 with $Ma=6$, $Re=6.37\times 10^6$, $\alpha$ = 8°, $T_{\infty}=58.54$ K, and $T_w/T_0=0.61$ which is far beyond the initial parameter space ($Re=3.36\times 10^6 \sim 5.04\times 10^6$, $\alpha$ = 9°$\sim$ 18°, $T_{\infty}=100$ K $\sim 150$ K) by fine-tuned model C as an example.
\begin{figure}[htb!]
	\centering
	\includegraphics[trim = 4cm 5.5cm 4cm 4cm, clip,width=0.9\textwidth]{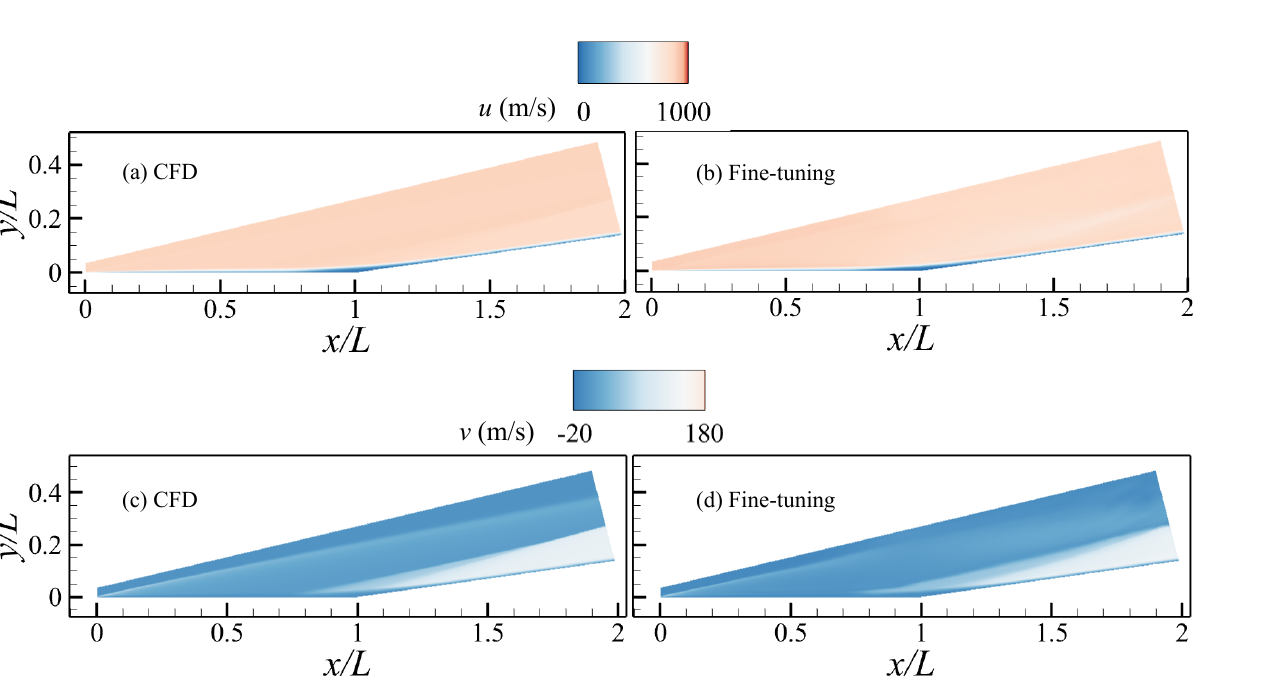}
	\caption{{Comparison of $u$ and $v$ velocity fields of ground truth and fine-tuning results at $Ma=6$, $Re=6.37\times10^6$, $\alpha=8^\circ$, $ T_{\infty}$ = 58.54 K, and  $ T_w /T_0 $ = 0.61.}}
	\label{figs:block-sliding-compare}
\end{figure}
\begin{figure}[htb!]
	\centering
	\includegraphics[trim = 4cm 5.5cm 4cm 4cm, clip,width=0.9\textwidth]{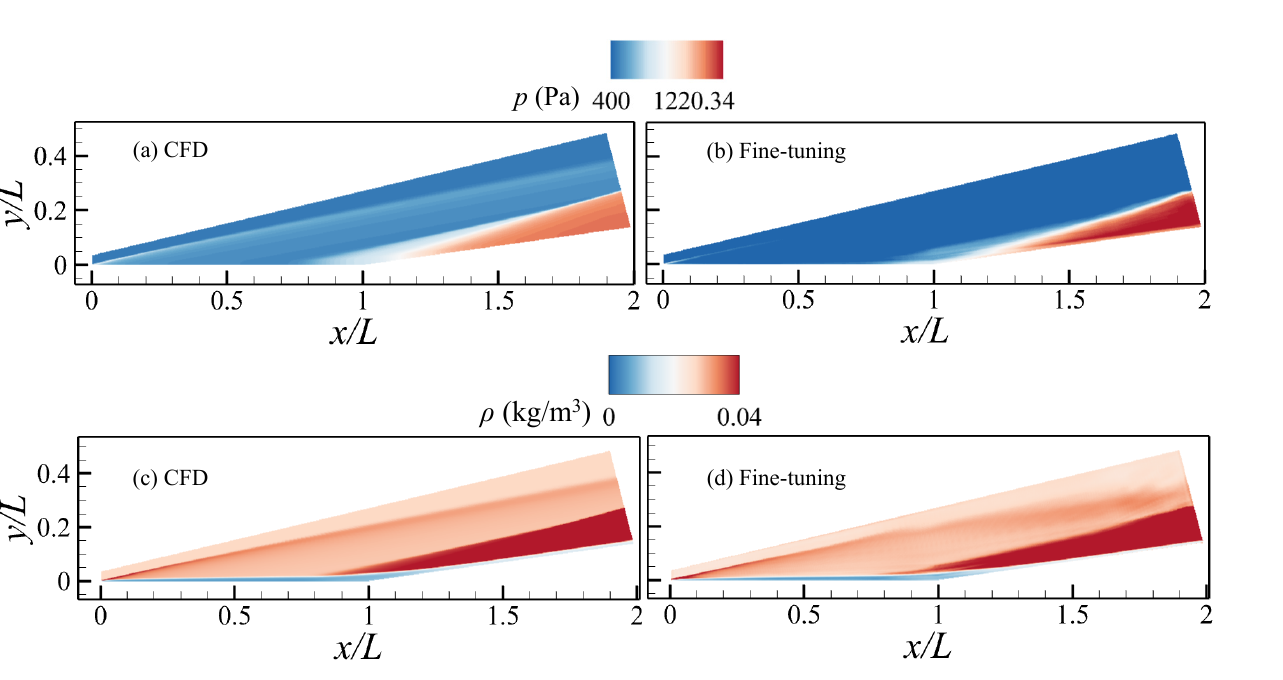}
	\caption{{Comparison of density and pressure fields of ground truth and fine-tuning model C at $Ma=6$, $Re=6.37\times10^6$, $\alpha=8^\circ$, $ T_{\infty}$ = 58.54 K, and  $ T_w /T_0 $ = 0.61.}}
	\label{figs:block-sliding-compare}
\end{figure}

As shown in Figures 19 and 20, which present the velocity, pressure, and density flow fields from both the ground truth and the fine-tuned model C, the proposed fine-tuning approach successfully predicts flow conditions far beyond the initial training parameter space and roughly shows good agreement with the reference data. It should be noted, however, that while the fine-tuned model captures the overall flow structures well, some discrepancies in local flow details are observed. These discrepancies can be primarily attributed to inherent differences between the experimental environment and the numerical setup used in the DNS. Factors such as surface roughness, freestream disturbances or minor geometric deviations from the idealized CAD model—difficult to fully avoid in experiments—can promote earlier boundary layer transition or alter shock wave-boundary layer interactions. Such effects are challenging to quantify and accurately replicate in the numerical setup, leading to systematic deviations in the predicted flow behavior.  

A similar trend is observed in the predictions of the surface pressure. For instance, Figure 21 compares the surface pressure coefficient obtained from the experiment, DNS and the predictions of both model D, the pre-trained and fine-tuned models for case 4 ($Ma=4$, $Re=7.22\times 10^6$, $\alpha$ = 10°, $T_{\infty}=70.7$ K and $T_w/T_0=0.99$).
\begin{figure}[htb!]
	\centering
	\includegraphics[trim = 2cm 1.2cm 3.5cm 2cm, clip,width=0.75\textwidth]{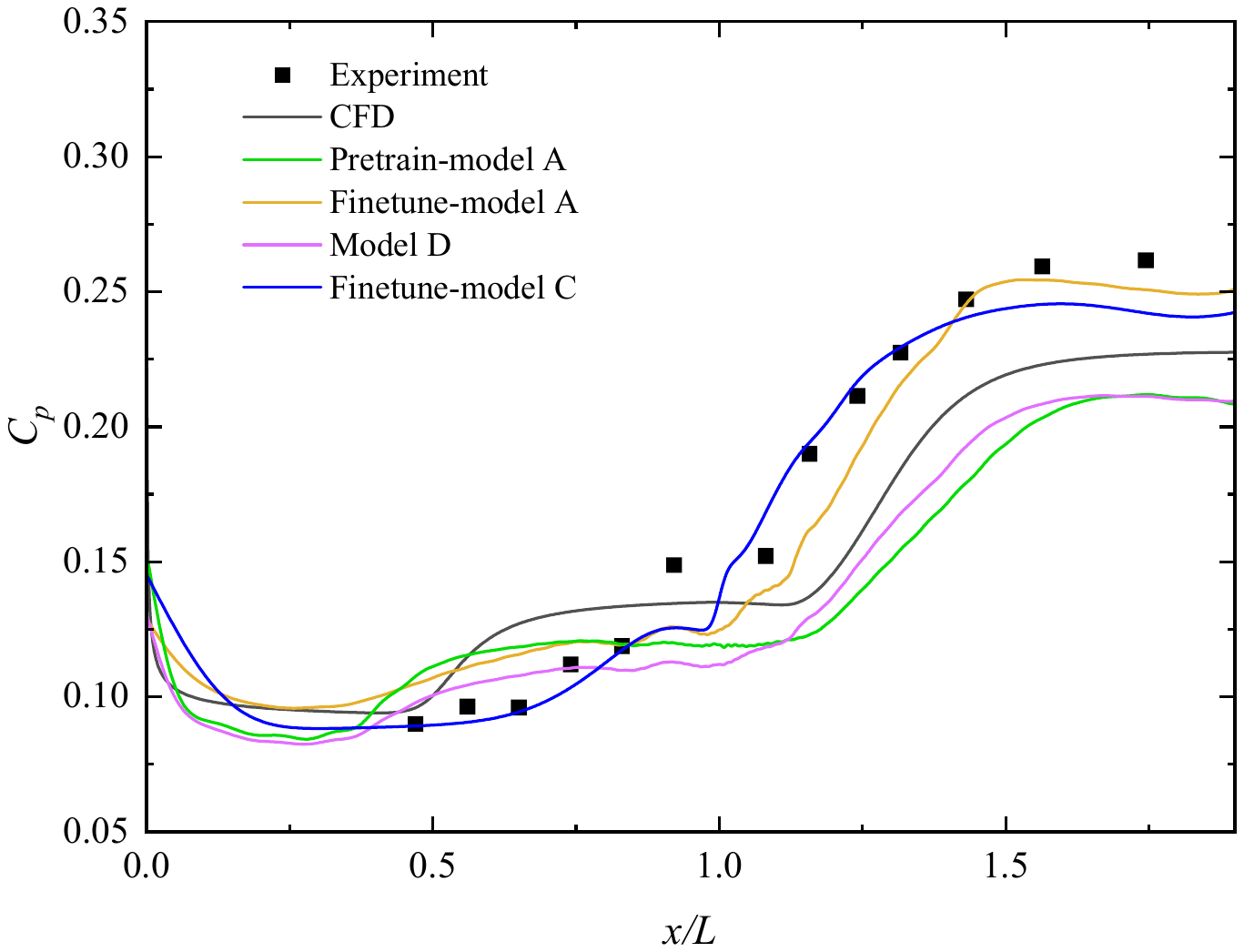}
	\caption{{Comparison of surface pressure coefficient of experiment, DNS and model predictions at $Ma=4$, $Re=7.22\times10^6$, $\alpha=10^\circ$, $ T_{\infty}$ = 70.7 K, and  $ T_w /T_0 $ = 0.99.}}
	\label{figs:block-sliding-compare}
\end{figure}
The results reveal discrepancies between experimental and DNS data when the ramp angle is large. Notably, the model fine-tuned by schlieren images shows improved agreement with the experimental surface pressure coefficient compared to models based on CFD-only (pre-trained model A) or hybrid CFD–experimental datasets (model D). This suggests that the fine-tuning process allows the originally ideal CFD-based model to incorporate real-world effects such as freestream disturbances, making its predictions more consistent with wind-tunnel measurements.

It is acknowledged that minor oscillations are present, especially in the predicted pressure coefficient curve of fine-tuned model A, which we attribute to discontinuities between adjacent patches \cite{Ma2025ContinuousViT}. This behavior is inherent in the ViT architecture. By dividing input images into a sequence of patches, the ViT model enhances its capacity for capturing global dependencies, although at the cost of reduced sensitivity to local details and spatial continuity. The self-attention mechanism enables direct interaction between any two patches, effectively building a global contextual representation. Nevertheless, this coarse, grid-based partitioning disrupts fine-grained pixel structures within patches and natural transitions across patch boundaries. As a result, the model inherently struggles to capture local features such as edges and near-wall flow structures \cite{9716741}.

In summary, the proposed fine-tuning framework effectively bridges the gap between ideal CFD models and real experimental conditions. By leveraging experimental schlieren images, the fine-tuned models incorporate physical effects such as freestream disturbances and surface roughness, which are challenging to simulate accurately in DNS. As a result, the predictions of the velocity, pressure and density fields will become more consistent with the real-world wind-tunnel. This approach demonstrates a viable pathway to enhance the physical fidelity of data-driven flow models using limited experimental data.

%
%
%
\section{Conclusion}
\label{sec1}
This study systematically investigates the prediction of hypersonic flow field over a compression ramp using neural networks, with a focus on the effective fusion of simulation and experimental data. The proposed methodology establishes a progressive framework: Initially, the fusion of experimental data with numerical simulations improves prediction accuracy; Subsequently, a pre-trained model is fine-tuned with experimental schlieren images to extend its applicability to real-world conditions. The main findings are summarized as follows:

(i) The integration of experimental data significantly enhances prediction accuracy. With the attention-based CNN model, the mean absolute error of pressure decreases from 0.012549 to 0.010962 (a reduction of 12.6\%), and that of density decreases from 0.004607 to 0.004266 (a reduction of 7.4\%). The attention-based CNN applied in this work is also shown to outperform the ViT in capturing fine-grained flow features.

(ii) Physical analysis based on the hybrid-trained attention CNN model confirms its capability to support engineering design. The predicted flow fields are successfully applied to global stability analysis, providing reliable physical insights while substantially reducing computational cost.

(iii) The pre-training and fine-tuning strategy effectively bridges simulation and experiment. By fine-tuning a CFD-pre-trained model with  schlieren images, the velocity and pressure fields under experimental conditions are accurately reconstructed, demonstrating strong transferability and practical potential for engineering applications.

In conclusion, this work validates a structured data-fusion approach for hypersonic flow modeling, highlighting the value of hybrid data training and transfer learning in improving predictive accuracy and expanding model generalization.



\section*{Conflict of Interest}

The authors have no conflicts to disclose.

\section*{Data Availability Statement}

The data that support the findings of this study are available from the corresponding author upon reasonable request. 

\section*{Declaration of generative AI and AI-assisted technologies in the writing process}

During the preparation of this work the authors used ChatGPT in order to improve language. After using this tool, the authors reviewed and edited the content as needed and take full responsibility for the content of the publication.

\section*{Appendix}

%
%

\bibliographystyle{elsarticle-num}
\bibliography{viscous-damping}
%
%
\end{document}